\documentclass[pdflatex,sn-basic,twocolumn]{sn-jnl}% Basic Springer Nature Reference Style/Chemistry Reference Style

\usepackage{graphicx}%
\usepackage{multirow}%
\usepackage{amsmath,amssymb,amsfonts}%
\fontsize{8}{10}
\usepackage{amsthm}%
\usepackage[scr=rsfs]{mathalpha}%
\usepackage[title]{appendix}%
\usepackage{xcolor}%
\usepackage{textcomp}%
\usepackage{manyfoot}%
\usepackage{booktabs}%
\usepackage{lmodern}%
\usepackage{algorithm}%
\usepackage{algorithmicx}%
\usepackage{algpseudocode}%
\usepackage{listings}%
\usepackage{siunitx}%
\usepackage{hyperref}%
\usepackage{natbib}
\usepackage{placeins}
\usepackage{longtable}

\usepackage{hyperref}
\usepackage{aas_macros}

\begin{document}

\title[Optical variability]{A comparative analysis of the long-term  optical variability characteristics of narrow and broad-line Seyfert 1 galaxies at $z$ $>$ 0.8}


\author*[1,2]{\fnm{Aratrika} \sur{Dey}}\email{aratrika@iiap.res.in}

\author[1,2]{\fnm{Stalin} \sur{C.S.}}\email{stalin@iiap.res.in}
%%\equalcont{These authors contributed equally to this work.}

\author[3,4]{\fnm{Akshith} \sur{S}}\email{akshithsatheesh@gmail.com}
%%\equalcont{These authors contributed equally to this work.}

\author[5]{\fnm{Suvendu} \sur{Rakshit}} \email{suvendu@aries.res.in}

\affil[1]{\orgname{Indian Institute of Astrophysics}, \orgaddress{\street{Block II, Koramangala}, \city{Bangalore}, \postcode{560034}, 
\state{Karnataka}, \country{India}}}

\affil[2]{\orgname{Pondicherry University}, \orgaddress{\street{R.V. Nagar, Kalapet}, \city{Puducherry}, \postcode{605014}, 
\state{Puducherry}, \country{India}}}

\affil[3]{\orgdiv{Department of Physics}, \orgname{St. Joseph's University}, \orgaddress{\street{36, Langford Road, Langford Gardens}, \city{Bangalore}, \postcode{560027}, 
\state{Karnataka}, \country{India}}}

\affil[4]{\orgname{Azim Premji University}, \orgaddress{\street{Burugunte Village, Survey No 66, Bikkanahalli Main Rd, Sarjapura,}, \city{Bangalore}, \postcode{562125}, \state{Karnataka}, \country{India}}}

\affil[5]{\orgname{Aryabhatta Research Institute of Observational Sciences}, \orgaddress{\street{Manora Peak}, \city{Nainital}, \postcode{263002}, 
\state{Uttarakhand}, \country{India}}}

%%\affil[3]{\orgdiv{Department}, \orgname{Organization}, \orgaddress{\street{Street}, \city{City}, \postcode{610101}, \state{State}, \country{Country}}}


\abstract{We present the results on a comparative analysis of the 
long-term optical variability characteristics of high-redshift narrow-line Seyfert 1 (NLSy1) galaxies and broad-line Seyfert 1 (BLSy1) galaxies. 
Our sample spanning the redshift range of $0.8 < z < 2.6$,  comprises 2490 NLSy1 and 2490 BLSy1 galaxies matched in the optical brightness$-$redshift plane.
We used V-band data from the  Catalina Real-Time Transient Survey covering 
a baseline of 5$-$9 years, and g, r, and i-band data from the Zwicky Transient Facility spanning 5 to 6 years. To characterise variability we estimated the amplitude of variability ($\sigma_m$). We found that NLSy1 galaxies generally exhibit lower $\sigma_m$ compared to BLSy1 galaxies. In both the NLSy1 and BLSy1 galaxy samples, we observed a wavelength dependent variability using data from Zwicky Transient Facility, with $\sigma_m$ gradually increasing towards shorter wavelengths. We found an anti-correlation  between $\sigma_m$ and Eddington ratio in the g, r, and i bands for BLSy1 galaxies, whereas this anti-correlation is observed only in the r and i bands  for NLSy1 galaxies. For both NLSy1 and BLSy1 galaxies, we found no correlation between $\sigma_m$ in g, r, and i bands and black hole mass. In both the samples, we found that $\sim$90\% of the sources showed a bluer when brighter trend. We found no significant time lag between variations in g and r bands in both BLSy1 and NLSy1 galaxies. The observed long-term trends in the optical light curves of our high redshift sample of BLSy1 and NLSy1 galaxies could be driven by variations in the accretion disk.}

\maketitle

\section{Introduction}\label{sec1}
Active galactic nuclei (AGN), powered by accretion of matter onto supermassive black holes (SMBHs) located at the centres of galaxies \citep{1984ARA&A..22..471R}, are among the most luminous extragalactic sources, that emit radiation across the electromagnetic spectrum with bolometric luminosities as large as 10$^{48}$ erg s$^{-1}$ \citep{2002ApJ...579..530W}.  Approximately 15\% of AGN emit 
strongly at radio wavelengths \citep{kellermann1989} with a majority of them exhibiting powerful, relativistic, Doppler-boosted jets. A defining characteristic of AGN, known since their discovery \citep{1963Natur.197.1041G} is their aperiodic continuum flux variability  across the entire accessible electromagnetic
spectrum. This variability is known to occur over a wide range of  amplitudes and timescales ranging from minutes to days \citep{1995ARA&A..33..163W,1997ARA&A..35..445U,2018Galax...6....2B,2021Galax...9..114W,2022A&A...668A.152P,2024ApJ...963...47P}.

Flux monitoring observations of AGN are important as they 
play a crucial role in helping us trace the dominant physical
processes in them \citep{2022ApJ...933...42A, 2023ApJS..265...51A} and
understand the nature of their central engines. Various theoretical models have been proposed to explain the observed flux variability over different timescales, including accretion disk instabilities, fluctuations in the mass accretion rate, multiple supernovae, micro-lensing events, Poisson processes, the damped random walk model and the damped harmonic oscillator model \citep{1997MNRAS.286..271A, 1998ApJ...504..671K,  2000A&AS..143..465H, 
2000ApJ...544..123C, 2009ApJ...698..895K, 2025ApJ...992..130Y}. 
Extensive optical photometric monitoring of large samples of AGN has revealed connections between the observed variability and several  physical parameters of the sources, such as the dependency of the variability amplitude with wavelength, luminosity, redshift, and black hole mass ($M_{BH}$) \citep[][and references therein]{1994MNRAS.268..305H, 1996ApJ...463..466D, 2004ApJ...601..692V, 2007MNRAS.375..989W, 2009ApJ...698..895K, 2009ApJ...696.1241B, 2010ApJ...721.1014M, 2011A&A...525A..37M}.
Despite these significant observational and theoretical efforts, the fundamental physical mechanisms driving AGN variability remain poorly understood. 

\begin{figure}[!htbp]
\includegraphics[scale=0.09]{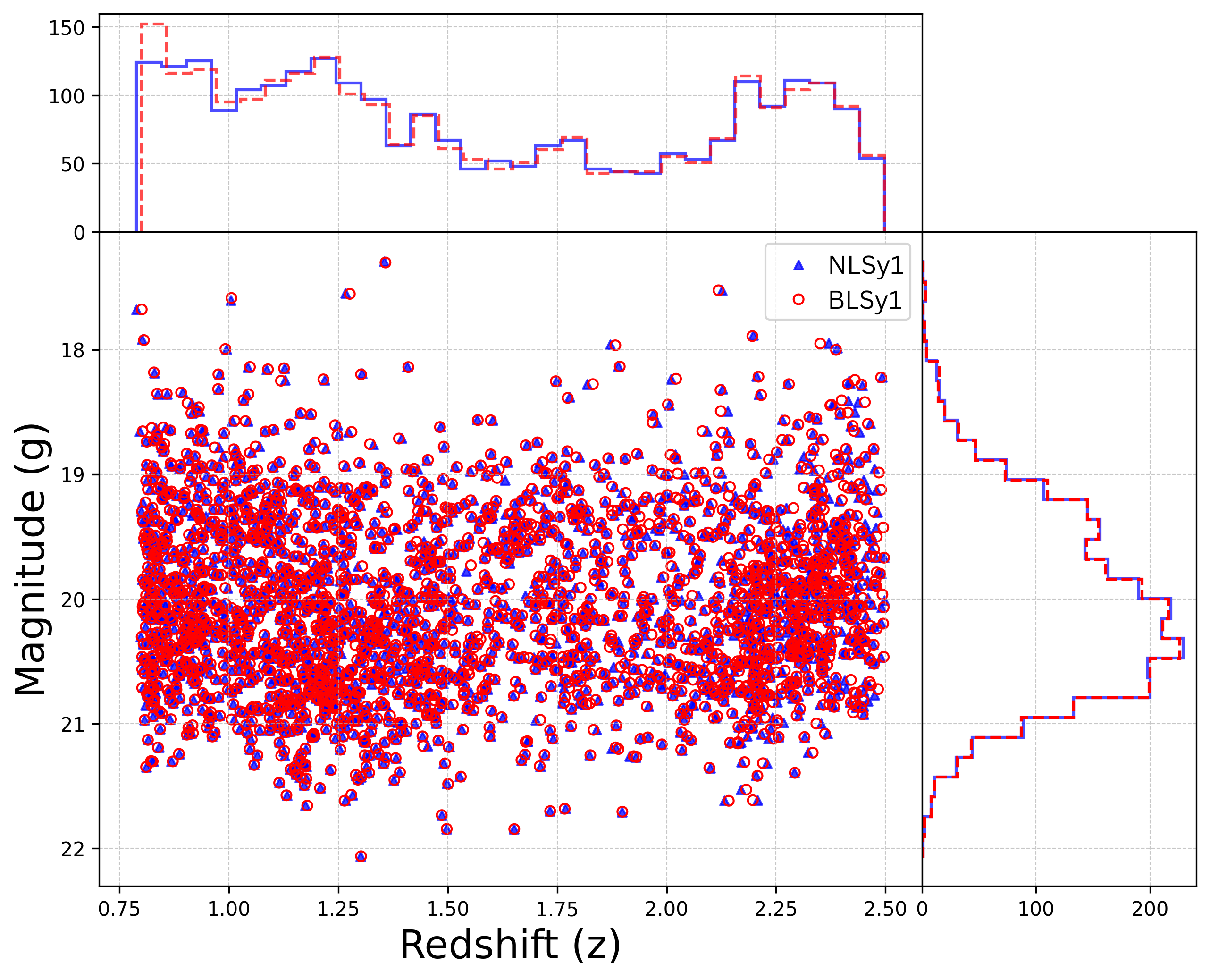}
\caption{Positions of the sources in the g-band magnitude and redshift plane. Here, blue filled triangles are NLSy1 galaxies and red open circles are BLSy1 galaxies. Also shown are the distributions of the redshifts (top panel) and g-band magnitudes (right panel) for NLSy1 galaxies (blue solid line) and BLSy1 galaxies (red dashed line) respectively.}
\label{fig-1}
\end{figure}

Among the various classes of AGN, narrow-line Seyfert 1 (NLSy1) galaxies, first identified about
four decades ago, represent a distinct subclass of Seyfert 1 galaxies.
They are characterized by the full width at half maximum (FWHM) of the
$\mathrm {H\beta ~line  < 2000\, km\,s^{-1}}$, significantly stronger Fe II emission
lines compared to broad-line Seyfert 1 (BLSy1) galaxies and flux ratio of [OIII]
to H$\beta$ less than 3 \citep{1985ApJ...297..166O, 1989ApJ...342..224G, 2001A&A...372..730V}.
Moreover, relative to their broad-line counterparts, NLSy1 galaxies also exhibit
more pronounced soft X-ray variability and steeper X-ray spectra \citep{1999ApJS..125..317L,
2004AJ....127.1799G}. They are associated with lower black hole masses (10$^6$ $-$ 10$^8\, M_{\odot}$), and higher Eddington ratios \citep{2012AJ....143...83X}. In contrast, BLSy1 galaxies generally
host more massive black holes ($\gtrsim$ 10$^8 \, M_{\odot}$; 
\citealt{2006ApJS..166..128Z,2012AJ....143...83X}). However, some studies
suggest that NLSy1 galaxies may harbour SMBHs with masses comparable to those
in BLSy1 galaxies and even blazars (\citealt{2013MNRAS.431..210C,2016MNRAS.458L..69B,2019ApJ...881L..24V}).

NLSy1 galaxies have been studied for their optical flux variability characteristics.
Early investigations primarily focused on short-term flux variations on
nightly and daily timescales \citep{1999MNRAS.304L..46Y,2000NewAR..44..539M,
1999MNRAS.304L..46Y,2000NewAR..44..539M, 2006ARep...50..708D}.
For instance, \citet{2006ARep...50..708D} conducted a detailed study of the NLSy1 galaxy
Ark 564, reporting a low variability amplitude of 0.1$-$0.2 mag. Similarly,
\cite{2004ApJ...609...69K} based on a study of six NLSy1 galaxies, concluded that NLSy1 galaxies generally
exhibit lower variability compared to BLSy1 galaxies. However, both studies
were limited by sample sizes. Taking advantage of the multi-wavelength, multi-epoch
photometric data from the Sloan Digital Sky Survey (SDSS) and the NLSy1
galaxy catalog compiled by \cite{2006ApJS..166..128Z}, \cite{2010ApJ...716L..31A,2013AJ....145...90A} 
carried out a comparative analysis of the long-term optical
variability of a moderate sample of 55 NLSy1 and 108 BLSy1 galaxies.
They found that NLSy1 galaxies consistently show lower amplitude of variability compared to BLSy1 galaxies. Expanding the work of \cite{2010ApJ...716L..31A} on a larger sample of NLSy1 galaxies and  a matched sample of BLSy1 galaxies, \cite{2017ApJ...842...96R}, found that NLSy1 galaxies exhibit lower long-term optical variability compared to BLSy1 galaxies. More recently, using data from the Panoramic Survey Telescope and Rapid Response System (Pan-STARRS), \cite{2023Ap&SS.368...68W} carried out a comparative analysis of the long-term optical variability characteristics of NLSy1 and BLSy1 galaxies at $z$ $<$ 0.8. Their results also
support the findings of NLSy1 galaxies showing lower variability amplitude compared to BLSy1 galaxies. 

In addition to the long-term optical flux variations,  NLSy1 galaxies are also known to show optical variations within a night \citep{2013MNRAS.428.2450P,2017MNRAS.466.2679K,2022MNRAS.514.5607O}. All previous studies on the long-term optical flux
variations of NLSy1 galaxies have been limited to sources with $z < 0.8$. This limitation was due to the non-availability of high redshift NLSy1 galaxies, as NLSy1 galaxies are selected based on the width of the $H\beta$ emission line which lies in the optical 
region only for $z$ $<$ 0.8. To overcome this, \cite{2021ApJS..253...28R}, used the correlation observed between the FWHM of  MgII and $H\beta$ emission lines, and arrived at a sample of high redshift NLSy1 galaxies. Motivated by the availability
of this high-$z$ NLSy1 galaxies sample at $z > 0.8$, in this work we aim to
(i) characterise the long-term optical variability of NLSy1 galaxies at $z$ $>$ 0.8,  and (ii) compare their 
variability properties with those of their broad-line counterparts, the BLSy1 galaxies.  

The paper is organized as follows. The sample and data used in this work are detailed in 
Section 2. The analysis of the light curves is given in Section 3. Section 4 presents the results followed by the conclusions in Section 5.

\FloatBarrier

\begin{figure*}[!tp]
\vbox{
\hspace*{-1.5cm}\includegraphics[scale=0.5]{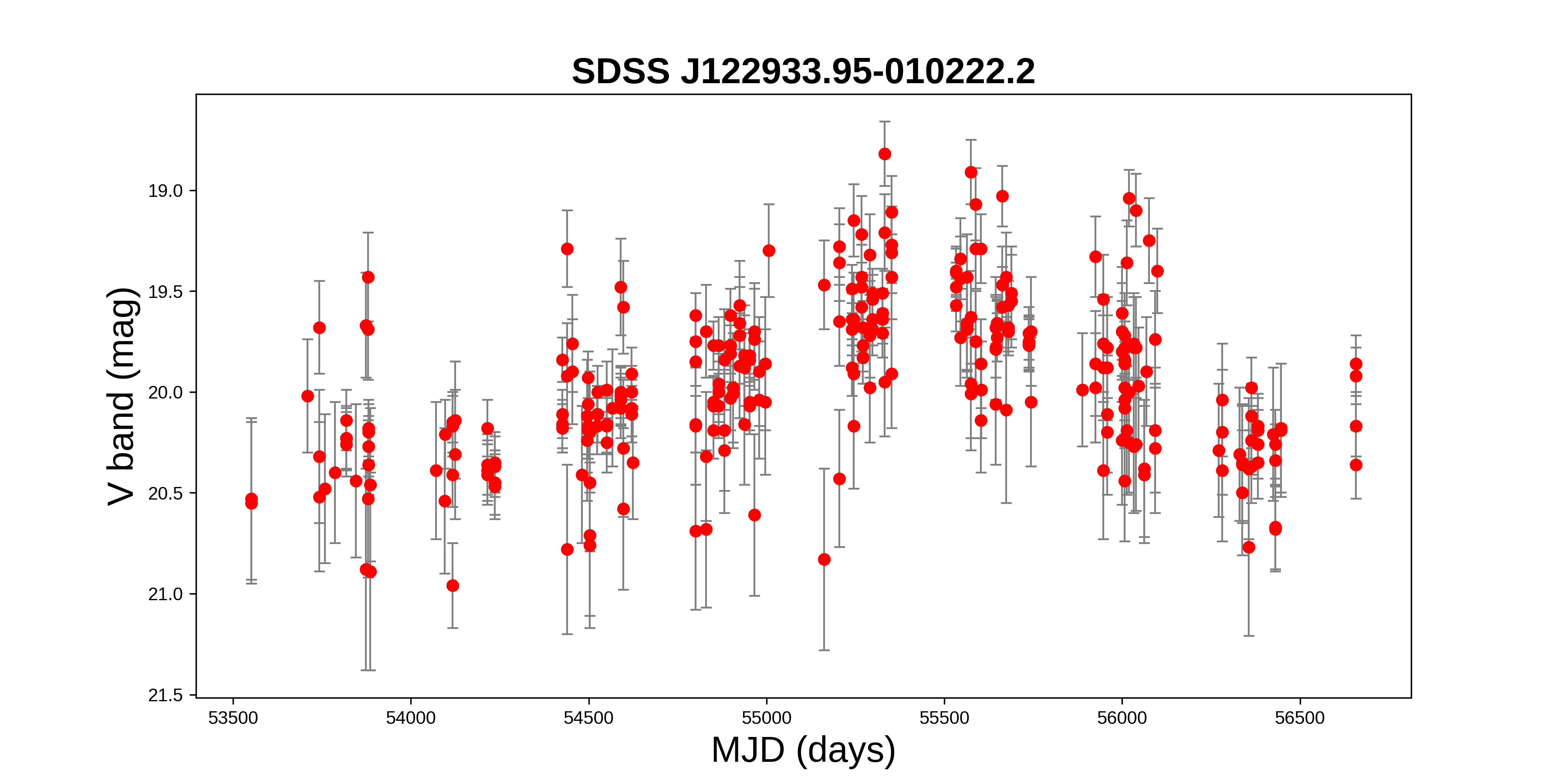}
\hspace*{-1.5cm}\includegraphics[scale=0.5]{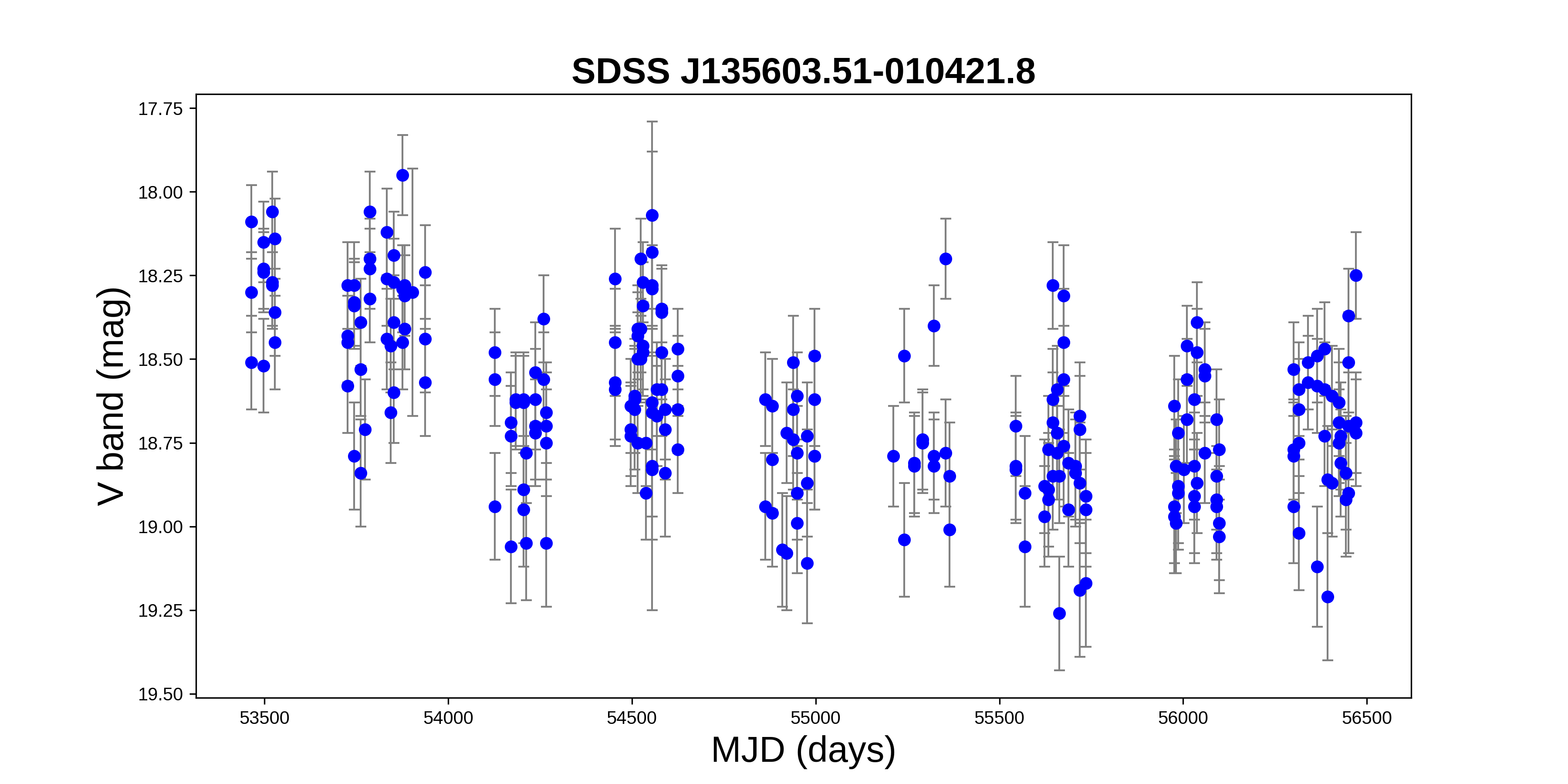}
     }
\caption{The V-band light curve from CRTS for the source SDSS J122933.95$-$010222.2 (top panel) and SDSS J135603.51$-$010421.8 (bottom panel).}
\label{fig-2}
\end{figure*}

\begin{figure*}[!tp]
\vbox{
\hspace*{-1.5cm}\includegraphics[scale=0.5]{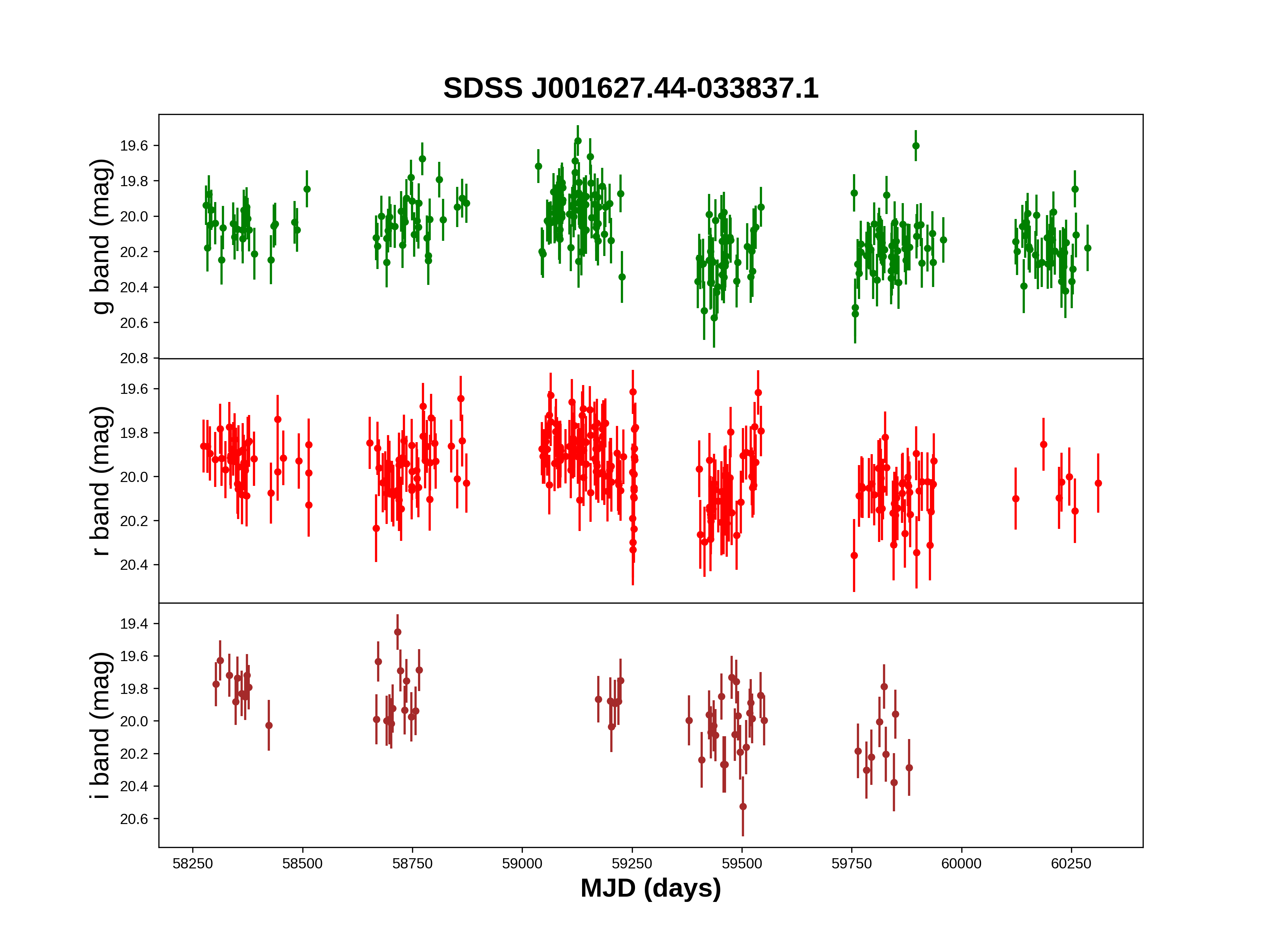}
\hspace*{-1.5cm}\includegraphics[scale=0.5]{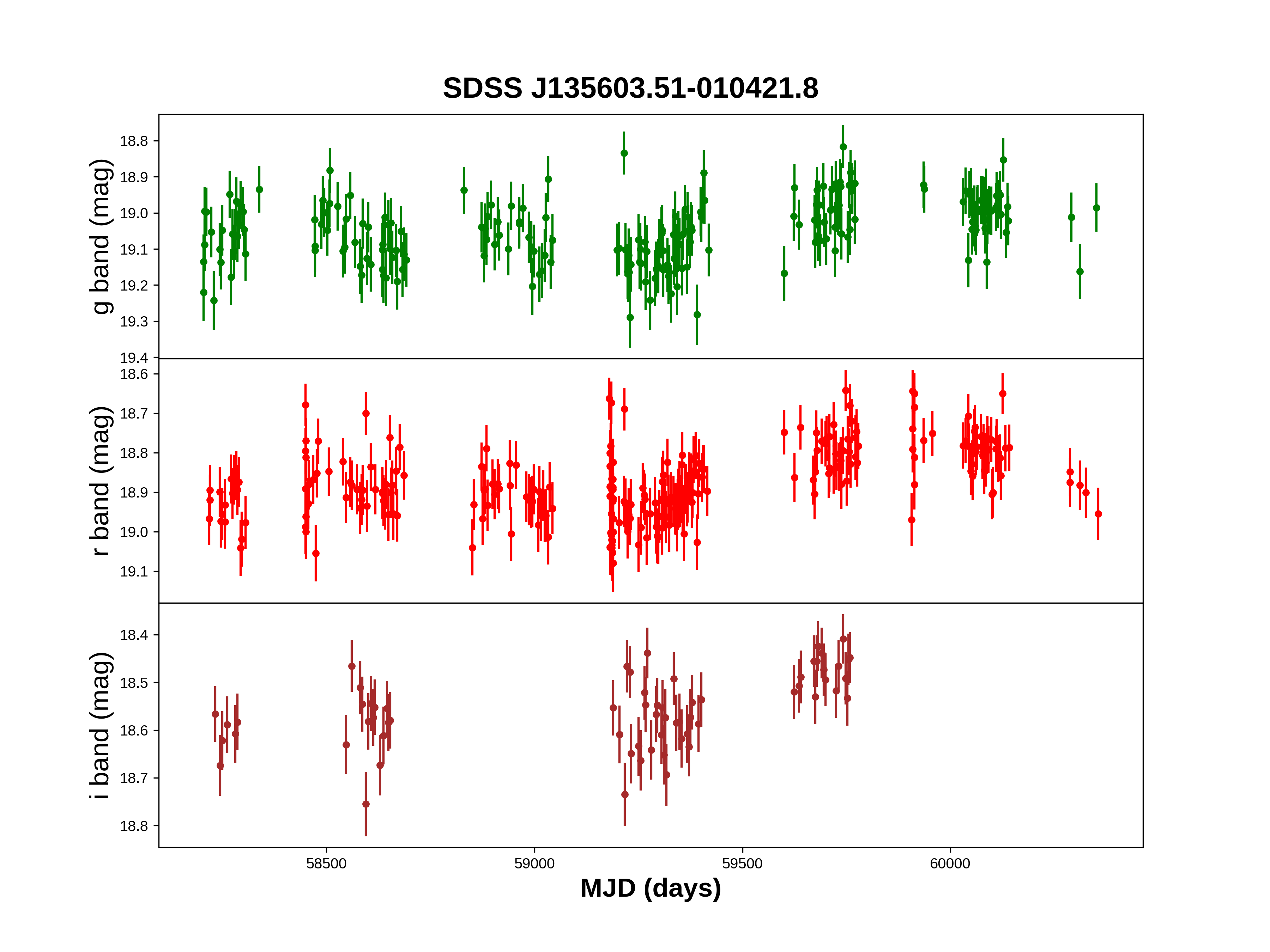}
     }
\caption{The light curves in g, r, and i bands from ZTF for the source SDSS J001627.44$-$033837.1 (top panel) and SDSS J135603.51$-$010421.8 (bottom panel).}
\label{fig-3}
\end{figure*}

\section{Sample and data}
\subsection{Sample}
Our sample consists of both NLSy1 and BLSy1 galaxies. According to the original definition, NLSy1 galaxies are characterised by the FWHM of their H$\beta$ line $<$ 2000 km s$^{-1}$ and a flux ratio of  [OIII] to H$\beta$ $<$ 3. However, applying these criteria restricts the 
identification of NLSy1 galaxies to redshifts $z < 0.8$, beyond which the H$\beta$ line
is no longer accessible in optical spectra. Consequently, for high redshift NLSy1 galaxies, 
the FWHM(H$\beta$) $<$ 2000 km s$^{-1}$ criterion cannot be directly applied. 
\cite{rakshit2020spectral} found a close correlation between the line widths of Mg II and H$\beta$, enabling the use of Mg II as a proxy for H$\beta$ at higher redshifts.   
Building on this, \cite{2021ApJS..253...28R} identified high-$z$ NLSy1 galaxies 
using the criterion FWHM(Mg II) $<$ 2000 km s$^{-1}$. Using this approach they 
compiled  a sample of 2684 high redshift NLSy1 galaxies, which forms the basis of our initial sample of NLSy1 galaxies.
To construct a control sample of BLSy1 galaxies, we used
the catalog of \citet{rakshit2020spectral}. This catalog contains measurements of the spectral properties for a total of 526,265 quasars selected from the fourteenth data release of the Sloan Digital Sky Survey quasar catalog. The primary objective of this  
work is to carry out a comparative analysis of the long-term optical variability characteristics of high redshift NLSy1 and BLSy1 galaxies, which requires a matched sample of NLSy1 and BLSy1 
galaxies. We therefore cross-matched both the catalogs, 
with the condition that for each NLSy1 galaxy in \citet{2021ApJS..253...28R} 
the BLSy1 galaxy (with M$_B$ $<$ $-$23 mag) in \citet{rakshit2020spectral} must match
within a magnitude difference of $\Delta$g = 0.01 mag and 
redshift difference of $\Delta$z = 0.05. Applying these criteria
we arrived at a final sample of 2490 NLSy1 galaxies and 2490 BLSy1 galaxies.
 The details on the full sample of both NLSy1 and BLSy1 galaxies used in this work are given in Tables \ref{table-1} and \ref{table-2} respectively.
The distributions of the two samples of sources in the magnitude-redshift plane are shown in Fig. $\ref{fig-1}$. From the two-dimensional Kolmogorov-Smirnov (KS) test  \citep{1992nrfa.book.....P}, we found a D-value of 0.0048 and a p-value of 0.8. The two samples are thus statistically indistinguishable.

\subsection{Photometric data}
For long-term optical monitoring data, we used the optical V-band
data from the Catalina Real-time Transient Survey (CRTS; \citealt{drake2009first}) 
and data in g, r, and i bands from the Zwicky Transient Facility (ZTF; \citealt{2019PASP..131a8002B}), Data Release 23. CRTS comprises three telescopes, namely, the $0.7$ m Catalina Schmidt Telescope with a field of view (FoV) of 8.1 deg$^{2}$, the $1.5$ m Mount Lemmon Survey reflector telescope located north of Tucson, Arizona
with a FoV of 1.2 deg$^{2}$, and the 0.5 m Uppsala Schmidt telescope at Siding Spring, Australia with a FoV of 4.2 deg$^{2}$. Each telescope has a 4k $\times$ 4k CCD and avoids the Galactic plane by $10 ~\deg$ because of high stellar crowding and patchy extinction. On a typical night, CRTS covers approximately 2500 deg$^{2}$ of the sky, taking 4 exposures per visit that have a temporal gap of 10 minutes. In CRTS, the images were acquired without using any filter \citep{2021MNRAS.507.4983S}. All data were processed in real-time using automated pipelines and calibrated to the Johnson V band. Observations were made over 21 nights per lunar cycle, reaching a V-band depth of around 19$-$20 mag. The accuracy of the photometry is limited by the transformation to the V band from unfiltered images.  

Similarly, the ZTF is a robotic time-domain survey with the  
48 inch Schmidt Telescope  at the Palomar Observatory \citep{2019PASP..131a8003M}.
It provides photometric observations in g, r, and i bands. The ZTF survey scans the northern sky at around 3 day cadence and the Galactic plane with a cadence of 1 night \citep{2025ApJ...992..153N}.
We cross-matched our sample of 2490 BLSy1 galaxies and 2490 NLSy1 galaxies with
both the CRTS and ZTF databases. For this, we used a search radius of 
2$^{\prime\prime}$. We also imposed a condition of each of the sources to have a minimum of 10 epochs of observations. For ZTF, among the quadrants, we chose a single
quadrant that has the maximum number of data points.  Also, we imposed an additional constraint for the light curves to have the ZTF catflag parameter to be zero \citep{2025MNRAS.543..121S}, which indicates that the data are good and free from any known quality issues\footnote{\url{https://irsa.ipac.caltech.edu/data/ZTF/docs/releases/dr23/ztf_release_notes_dr23.pdf}}. 
Once the light curves were generated, we inspected them for the presence of any outlier points in flux as well as points with large error bars. To eliminate such outliers, we applied
an iterative 3$\sigma$ clipping process, where points that deviate by more than three times the standard deviation from the mean of the light curves were discarded.
Sample CRTS V-band light curves for two sources are shown in Fig. $\ref{fig-2}$. 
Similarly,  sample light curves for two sources in g, r, and i bands from ZTF are shown in Fig. $\ref{fig-3}$.  

\FloatBarrier
\clearpage

\begin{table*}[!tp]
\small
\caption{Details of NLSy1 galaxies used in this work. For illustration, only 10 sources are listed. The full table is available in the electronic version of the article. The redshift ($z$), log (M$_{BH}$), the g-band and r-band magnitudes, and RA and Dec in degrees are given.}
\label{table-1}
\begin{tabular}{crrcccc} \hline
SDSS Name  & RA (2000)  & Dec (2000)  & $z$  & log (M$_{BH}$)  & g band  &r band \\ \hline
132434.41$-$010436.7 & 201.143375746727 & $-$1.076878850381  & 2.430 & 9.017 & 19.574 & 19.494  \\
135603.51$-$010421.8 & 209.014633535612 & $-$1.072744956619  & 1.932 & 8.940 & 18.837 & 18.742  \\
140932.85$-$002337.7 & 212.386889300289 & $-$0.393828509175  & 2.230 & 9.177 & 19.928 & 19.592  \\
141015.36$-$001418.9 & 212.564014288792 & $-$0.238589633151  & 1.876 & 8.199 & 19.290 & 19.037  \\
153722.75$-$001412.9 & 234.344802206704 & $-$0.236934341618  & 2.288 & 8.116 & 19.895 & 19.766  \\
144706.45+003246.1   & 221.776900236479 & 0.546149656977    & 1.158 & 7.758 & 19.659 & 19.436  \\
154817.13+003109.2   & 237.071396286412 & 0.519230751997    & 1.042 & 7.945 & 19.167 & 18.818  \\
123808.38+005246.3   & 189.534935757790 & 0.879548005665    & 0.803 & 7.484 & 19.372 & 19.101  \\
130039.38+005512.5   & 195.164119278637 & 0.920156923308    & 2.446 & 9.302 & 20.738 & 21.026  \\
113213.02$-$005245.0 & 173.054279176851 & $-$0.879183506552 & 1.450 & 8.226 & 19.298 & 19.202  \\ \hline
\end{tabular}
\end{table*}

\begin{table*}[!tp]
\small
\caption{Details of BLSy1 galaxies used in this work. For illustration, only 10 sources are listed. The full table is available in the electronic version of the article. The redshift ($z$), log (M$_{BH}$), the g-band and r-band magnitudes, and RA and Dec in degrees are given.}
\label{table-2}
\begin{tabular}{crrcccc} \hline
SDSS Name  & RA (2000)  & Dec (2000)  & $z$  & log (M$_{BH}$) & g band  & r band \\ \hline
113548.26+195454.2 & 173.951104721395   &  19.9150719063512 & 2.423   & 8.671 & 19.573 & 19.642 \\
083342.44+422703.9 & 128.426848280797   &  42.4510965977839 & 1.936   & 8.636 & 18.835 & 18.531 \\
001627.44$-$033837.1 & 4.114373051468 & $-$3.64364075221369 & 2.240   & 9.146 & 19.930 & 19.839 \\
092106.20+592958.9 & 140.275849405667   &  59.4997058377571 & 2.289   & 9.141 & 19.898 & 19.955 \\
142251.80+411918.6 & 215.715832327602   &  41.3218391570484 & 1.159   & 8.695 & 19.661 & 19.057 \\
104301.84+213425.0 & 160.757676193218   &  21.5736121910411 & 1.041   & 8.281 & 19.169 & 19.039 \\
130454.85+412742.9 & 196.228575614162   &  41.4619413663558 & 0.804   & 8.765 & 19.369 & 19.440 \\
085752.45+190119.4 & 134.468576736806   & 19.0220762098554  & 2.450   & 8.894 & 20.738 & 20.708 \\
034343.51$-$061233.3 & 55.931303893832  & $-6$.20927702008073 & 1.457   & 9.024 & 19.295 & 19.178 \\
215313.78+241413.2 & 328.307429717364   &  24.2370092423323 & 1.559   & 8.650 & 19.207 & 19.027 \\ \hline
\end{tabular}
\end{table*}

\section{Light curve analysis}
To characterise variability of our sample of sources, we calculated the intrinsic 
amplitude of variability  following  \cite{sesar2007exploring} as 

\begin{equation}
    \Delta = \sqrt{\frac{1}{N-1}\sum_{i=1}^{N}(m_{i}-<m>)^{2}}
\end{equation}
Here, N is the number of observed points in the light curve, $m_{i}$ is magnitude 
of the i$^{th}$ measurement, and $<m>$ is the weighted mean of all the measurements.

The amplitude of variability obtained in this manner (referred to as $\sigma_{m}$ elsewhere) is defined as 
\begin{equation}
    \sigma_{m} = 
    \begin{cases}
      \sqrt{\Delta^{2}-\epsilon^{2}} & \text{if $\Delta>\epsilon$} \\
      0 & \text{otherwise}
    \end{cases}
    \label{eq:1}
\end{equation}
Here, $\epsilon$ is 
the mean error calculated from the errors in individual magnitude measurements 
$\epsilon_{i}$
as 
\begin{equation}
        \epsilon^{2} =\frac{1}{N}\sum_{i=1}^{N}{\epsilon_{i}}^{2}
        \label{eq:2}
\end{equation}

\begin{table*}[!tp]
\small
\caption{Results on the variability analysis of the sample of NLSy1 and BLSy1 galaxies. Here, N refers to the number of sources and {A($\sigma_m$) and M($\sigma_m$) are the mean and median values of $\sigma_m$ (in mag) }respectively} 
\label{table-3}

\hspace*{5.0cm} NLSy1 galaxies

\vspace*{0.2cm}
\resizebox{\textwidth}{!}{%
\begin{tabular}{lccccccccc} \hline
        &\multicolumn{3}{c}{Total Sample}  &\multicolumn{3}{c}{Radio-detected} &\multicolumn{3}{c}{Radio-undetected} \\
Filter  & N  & A($\sigma_m$)            & M($\sigma_m$) & N & A($\sigma_m$)    & M($\sigma_m$)   & N  & A($\sigma_m$) & M($\sigma_m$)\\ \hline
V     & 2048 & 0.28 $\pm$ 0.14 & 0.27   & 107   & 0.18 $\pm$ 0.12 & 0.18       & 2006  & 0.27 $\pm$ 0.14 & 0.27 \\
g     & 2040 & 0.10 $\pm$ 0.05 & 0.09   & 95    & 0.09 $\pm$ 0.04 & 0.09       & 1945  & 0.10 $\pm$ 0.05 & 0.09 \\
r     & 1964 & 0.08 $\pm$ 0.04 & 0.08   & 93    & 0.08 $\pm$ 0.04 & 0.07       & 1871  & 0.08 $\pm$ 0.05 & 0.08 \\
i     & 1507 & 0.07 $\pm$ 0.04 & 0.07   & 70    & 0.07 $\pm$ 0.05 & 0.06       & 1437  & 0.07 $\pm$ 0.04 & 0.07 \\ \hline
\end{tabular}%
}

\vspace*{0.2cm}
\hspace*{5.0cm} BLSy1 galaxies
\vspace*{0.2cm}

\resizebox{\textwidth}{!}{%
\begin{tabular}{lccccccccc} \hline
      &\multicolumn{3}{c}{Total Sample}  &\multicolumn{3}{c}{Radio-detected} &\multicolumn{3}{c}{Radio-undetected} \\
Filter  & N  & A($\sigma_m$)            & M($\sigma_m$) & N & A($\sigma_m$)    & M($\sigma_m$)   & N  & A($\sigma_m$) & M($\sigma_m$)\\ \hline
V     & 2042 & 0.29 $\pm$ 0.13 & 0.29   &  82 & 0.24 $\pm$ 0.12 & 0.23      & 1998  & 0.29 $\pm$ 0.14 & 0.29 \\
g     & 2149 & 0.13 $\pm$ 0.06 & 0.12   &  78 & 0.13 $\pm$ 0.07 & 0.11      & 2071  & 0.13 $\pm$ 0.06 & 0.12 \\
r     & 2127 & 0.10 $\pm$ 0.06 & 0.10   &  74 & 0.11 $\pm$ 0.07 & 0.10      & 2053  & 0.10 $\pm$ 0.06 & 0.09 \\
i     & 1524 & 0.09 $\pm$ 0.05 & 0.08   &  57 & 0.09 $\pm$ 0.06 & 0.09      & 1467  & 0.09 $\pm$ 0.05 & 0.08 \\ \hline
\end{tabular}%
}

\end{table*}

We also quantified the amplitude of variability of our sample sources, by fitting the light curves of  the sources (in V, g, r, and i bands) using the damped random walk (DRW) model implemented in JAVELIN\footnote{https://github.com/legolason/javelin-1.git}, a python code developed by \cite{2011ApJ...735...80Z}. The JAVELIN method has been widely used to model the continuum and emission-line light curves of AGN obtained as part of reverberation mapping experiments\citep{2012ApJ...744L...4G,2024ApJ...976..116P,2026JHEAp..5100552R}.

\FloatBarrier

\section{Results}
\subsection{Flux variability}
\subsubsection{CRTS}
\begin{figure*}[!tp]
\includegraphics[scale=0.37]{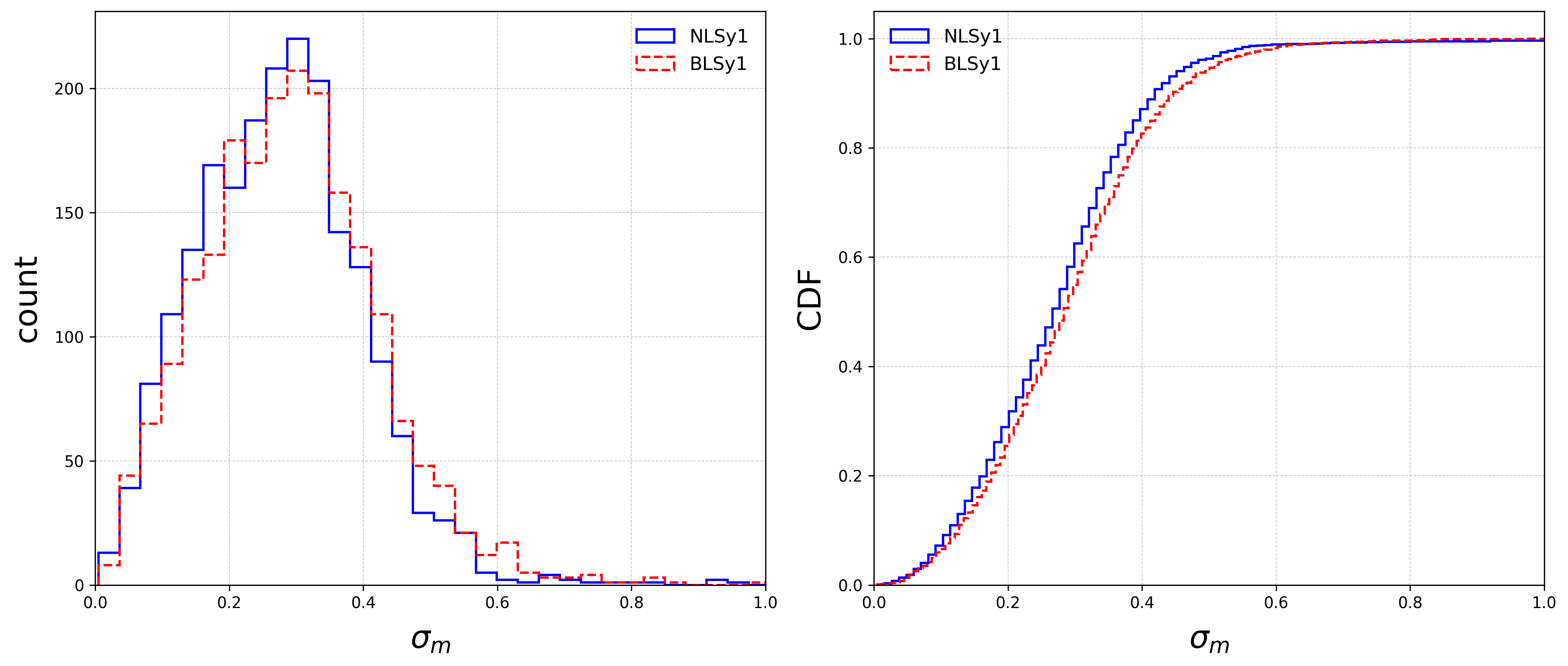 }
\caption{The distributions of variability amplitudes (left panel) and CDFs (right panel) from V-band observations of CRTS. The blue solid line is for NLSy1 galaxies and the red dashed line is for BLSy1 galaxies.}
\label{fig-4}
\end{figure*}

The results of flux variability for the NLSy1 and BLSy1 galaxy samples are given in Table \ref{table-3}. The histogram and the cumulative distribution function (CDF) of variability amplitudes in the V band based on data from CRTS are shown in Fig. $\ref{fig-4}$. For the full sample we found mean
$\sigma_m$ values of 0.28 $\pm$ 0.14 mag and 0.29 $\pm$ 0.13 mag for NLSy1 and BLSy1 galaxies respectively. Within uncertainties, the mean value of variability amplitude
found in NLSy1 and BLSy1 appears comparable. However, from  a two-sample KS test applied to the $\sigma_m$ distributions yielded a D-statistics value of 0.059  and a $p$-value of 0.002, indicating a statistically significant difference 
between the two distributions. We therefore conclude that, based on  CRTS V-band observations,  NLSy1 galaxies exhibit systematically lower variability amplitudes than BLSy1 galaxies. This trend is also clearly reflected in the CDFs shown in the right panel of Fig. $\ref{fig-4}$. The results
are in agreement with that known for low redshift NLSy1 
galaxies \citep{2017ApJ...842...96R,2023Ap&SS.368...68W}.

\subsubsection{ZTF}
\begin{figure*}[!tp]
\includegraphics[scale=0.12]{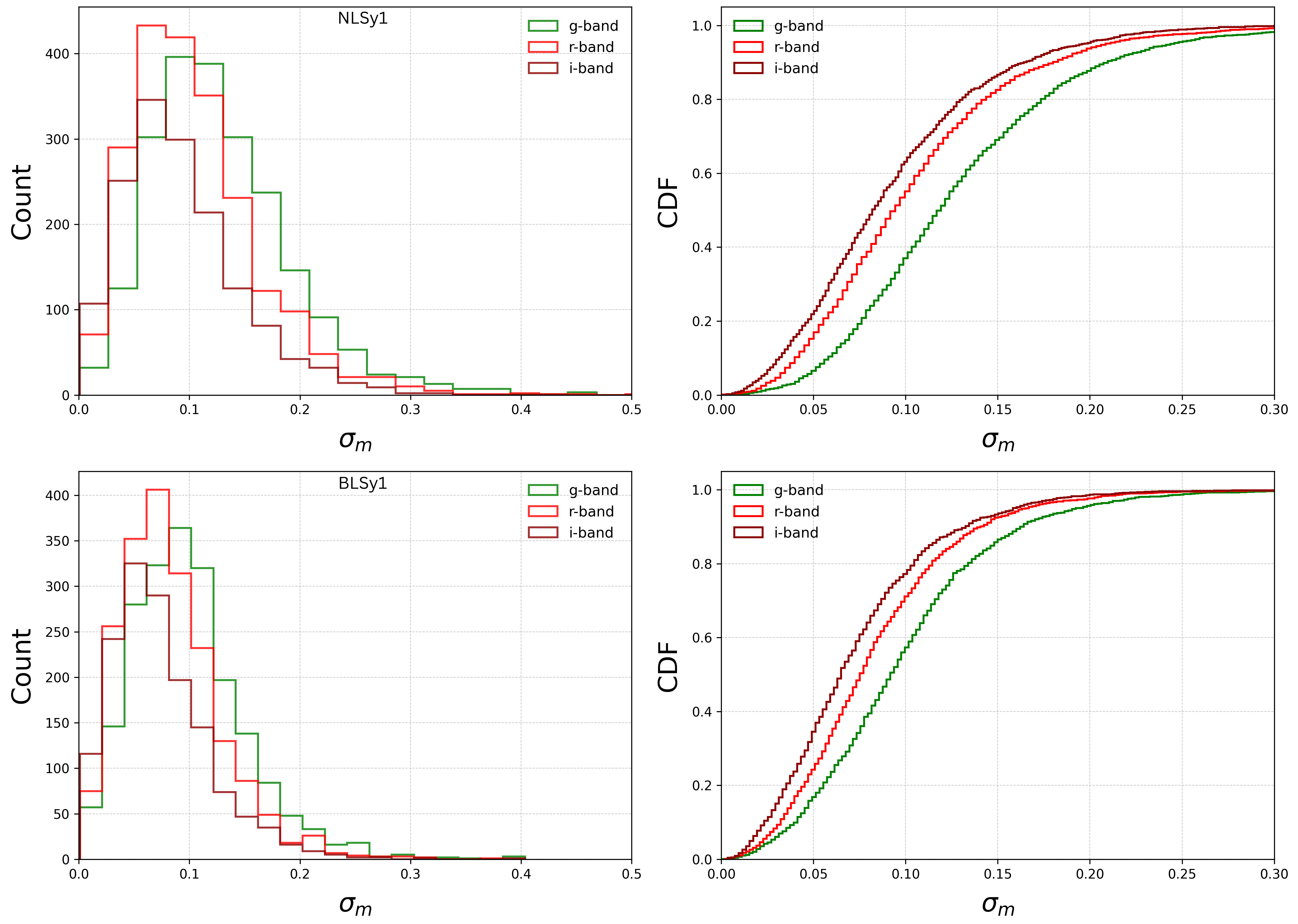}
\caption{The top left and top right panels show the distributions of $\sigma_m$ and CDFs in g (green), r (red), and i band (brown) from ZTF observations  of NLSy1 galaxies. Similarly, bottom left and bottom right panels show the distributions of $\sigma_m$ and CDF for BLSy1 galaxies in g (green), r (red), and i band (brown).} 
\label{fig-5}
\end{figure*}

We show in Fig. $\ref{fig-5}$ the histograms and CDFs of variability
amplitudes in g, r, and i bands for the combined sample of  both NLSy1 and BLSy1 galaxies. The results are given in Table \ref{table-3} for NLSy1 and BLSy1 galaxies. From Table \ref{table-3} it is evident that the mean $\sigma_m$ value in V band is larger than that of g, r, and i bands for both BLSy1 and NLSy1 galaxies. We note that the photometric errors computed by the CRTS pipeline are slightly underestimated. Therefore
the correction for the mean error in equation 2, is too small, making the deduced $\sigma_m$ larger \citep{2017MNRAS.472.4870S}. For the full sample of NLSy1 
galaxies, we found mean $\sigma_m$ values of 0.10 $\pm$ 0.05 mag, 0.08 $\pm$ 0.04 mag and 0.07 $\pm$ 0.04 mag in g, r, and i bands respectively. Similarly for the complete sample of BLSy1 galaxies we found mean $\sigma_m$ values in g, r, and i bands of 0.13 $\pm$ 0.06 mag, 0.10 $\pm$ 0.06 mag and 0.09 $\pm$ 0.05 mag respectively. 
Within error bars, the amplitude of variability between g, r, and i bands appear similar in both NLSy1 galaxies and BLSy1 galaxies. However, 
KS test indicates that the distribution of variability amplitudes between g, r, and i bands is different for both  NLSy1 and BLSy1 galaxies. In the case of NLSy1 galaxies, variability in g and r bands are greater than the variability in i band at the greater than 98\% confidence level, while variability in g band is greater than r band at greater than 90\% confidence. Similarly, in the case of BLSy1 galaxies, the amplitude of variations in g band is larger than that at i band at greater than 99\% confidence level, while no statistically significant differences could be ascertained between variations between g and r bands as well as between r and i bands.  The KS test supports a decreasing trend in variability amplitude from g to i band, confirming the wavelength dependence. This is also evident in the CDFs shown in Fig. $\ref{fig-5}$. The results of the KS test are shown in Table \ref{table-5}.

\begin{figure*}[!tp]
\includegraphics[scale=0.37]{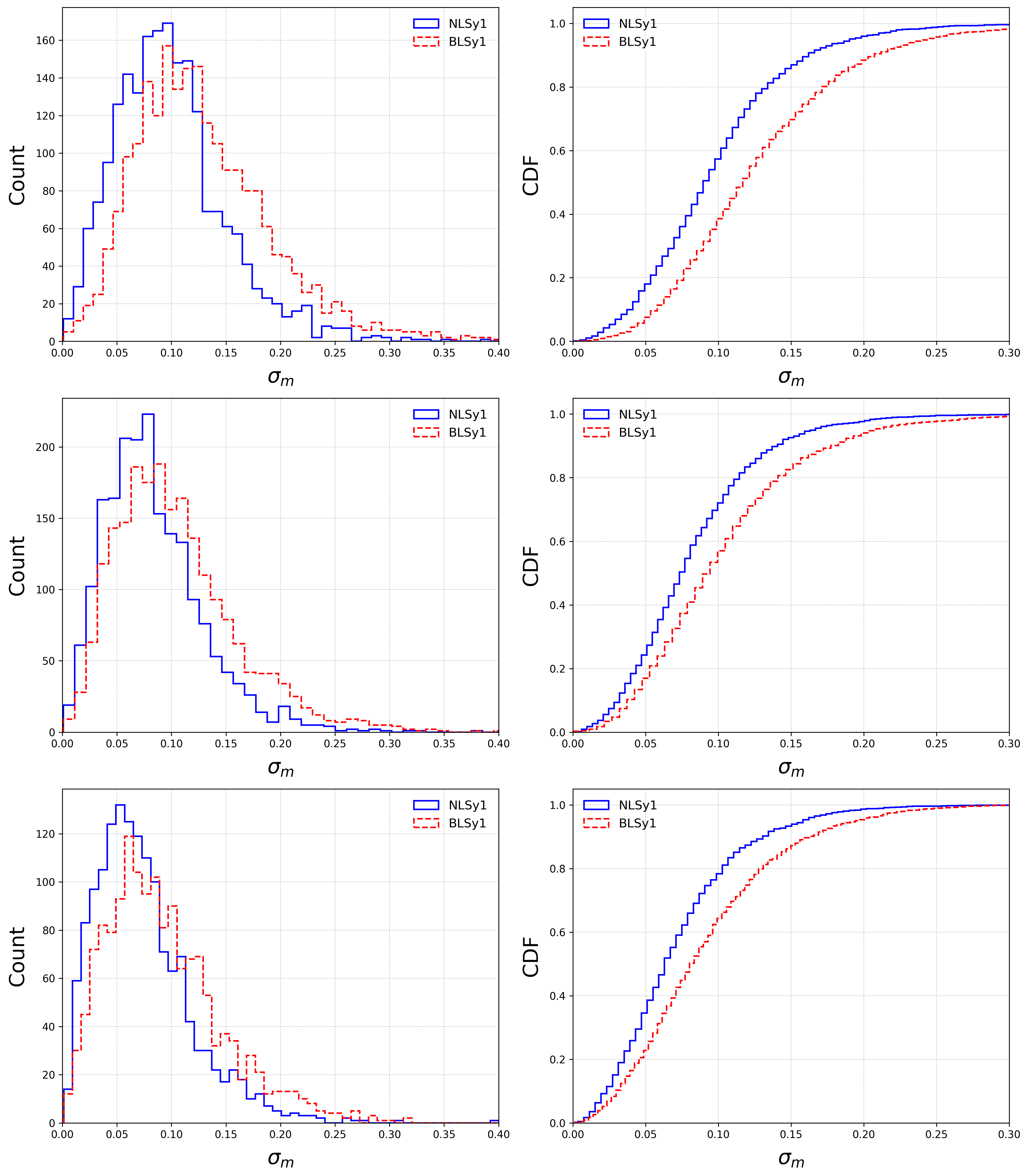}

\caption{Distributions of $\sigma_m$ and the CDFs in g band (top panels), r band (middle panels) and i band (bottom panels). The blue solid line is for NLSy1 galaxies and the red dotted line is for BLSy1 galaxies.}
\label{fig-6}
\end{figure*}

Similarly, within error bars, the mean variability amplitude in g, r, and i bands in NLSy1 galaxies are found to be similar to the mean variability
amplitude of BLSy1 galaxies. However, KS test indicates that the underlying distributions
are indeed different with BLSy1 galaxies more variable than NLSy1 galaxies in all the three bands 
at greater than 99\% confidence.
The histograms and CDFs of variability in g, r, and i bands between
BLSy1 and NLSy1 galaxies are shown in Fig. \ref{fig-6}  and
the results of the KS test are given in Table \ref{table-5}. 
In all the observed bands (V, g, r, and i) NLSy1 galaxies exhibit lower amplitude of variability than BLSy1 galaxies at greater than 99\% confidence level.
This reduced  variability in NLSy1 galaxies compared to BLSy1 galaxies can be understood in them having  higher Eddington ratio relative to BLSy1 galaxies. A higher Eddington ratio implies that the size of the emission region in NLSy1 galaxies is larger than that of BLSy1 galaxies leading to lower variability amplitude in them relative to BLSy1 galaxies \citep{2017ApJ...842...96R}.
These results indicate that (a) between NLSy1 and BLSy1 galaxies, NLSy1 galaxies tend to show lower variability amplitude than BLSy1 galaxies in all the wavebands investigated in this study, and (b) within NLSy1 and BLSy1 galaxies, there is a clear wavelength dependence, with the variability amplitude increasing towards shorter wavelengths.

\begin{figure*}[!tp]
\includegraphics[scale=0.1]{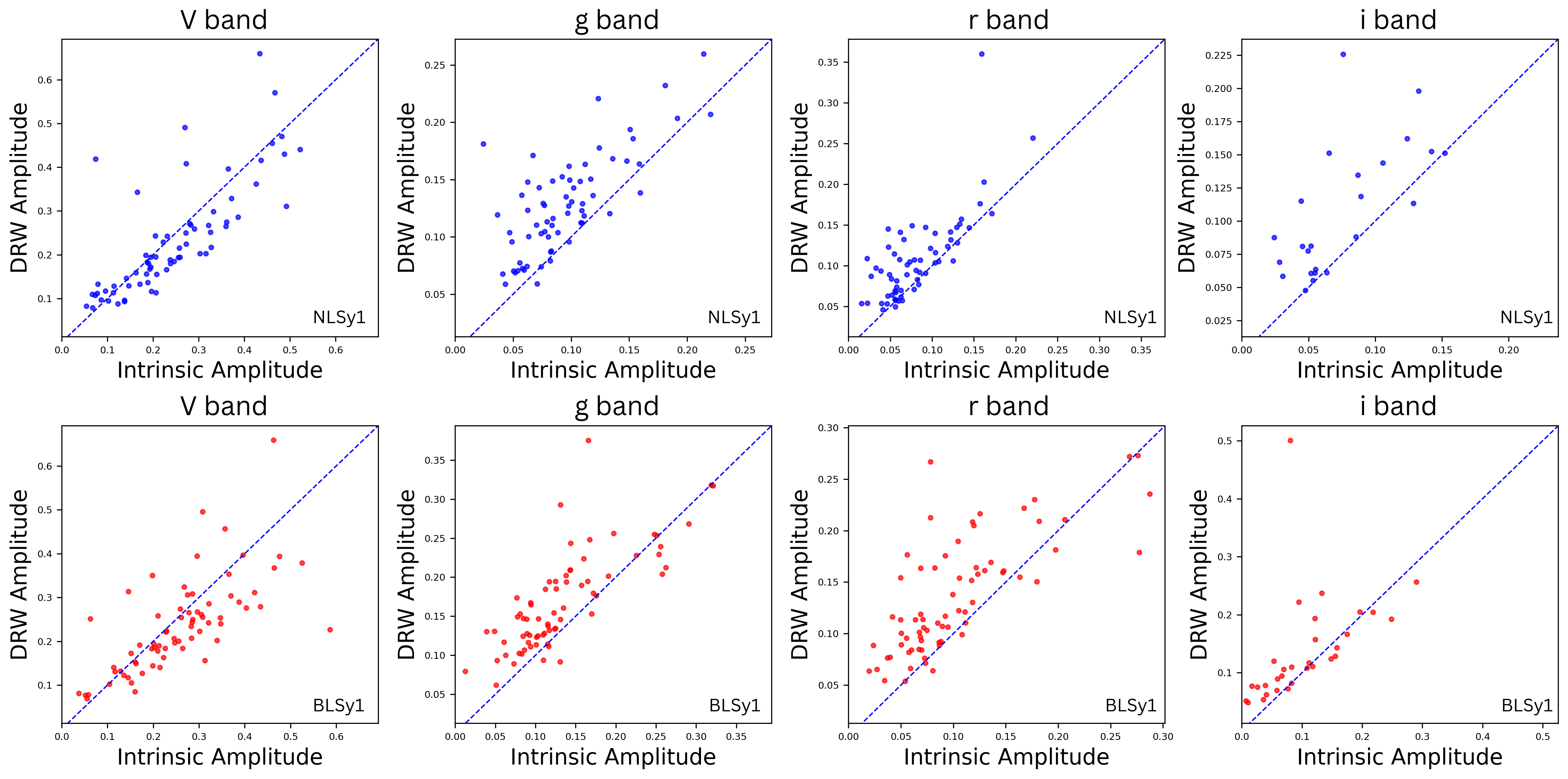}
\caption{Comparison between the two variability indicators used in this work, $\sigma_m$ and variability amplitudes obtained from DRW in different bands. The top panels are for NLSy1 galaxies and the bottom panels are for BLSy1 galaxies. The dotted lines represent the one to one correspondence between the two values.}
\label{fig-7}
\end{figure*}

We also compared the amplitudes of variability obtained from two variability measures, namely $\sigma_{m}$ obtained from the fitting of the light curves using JAVELIN and $\sigma_m$ estimated directly from the observed light curves. For this we randomly chose 100 sources, and show in Fig. $\ref{fig-7}$ the comparison between those two indicators of variability. The dashed lines
in Fig. $\ref{fig-7}$ show the one-to-one correspondence between the two values. The statistics of these two measurements on those 100 sources are given in Table \ref{table-4}. We found that the amplitudes of  variability found from JAVELIN are in agreement with that obtained directly from light curves, such that (i) NLSy1 galaxies are less variable than BLSy1 galaxies and (ii) the amplitude of variability gradually increases towards shorter wavelengths in both NLSy1 and BLSy1 galaxies. However, the amplitudes of variability from JAVELIN tend to be slightly larger than $\sigma_m$ in all the bands, except the V band.

\begin{figure*}[!tp]
\includegraphics[scale=0.1]{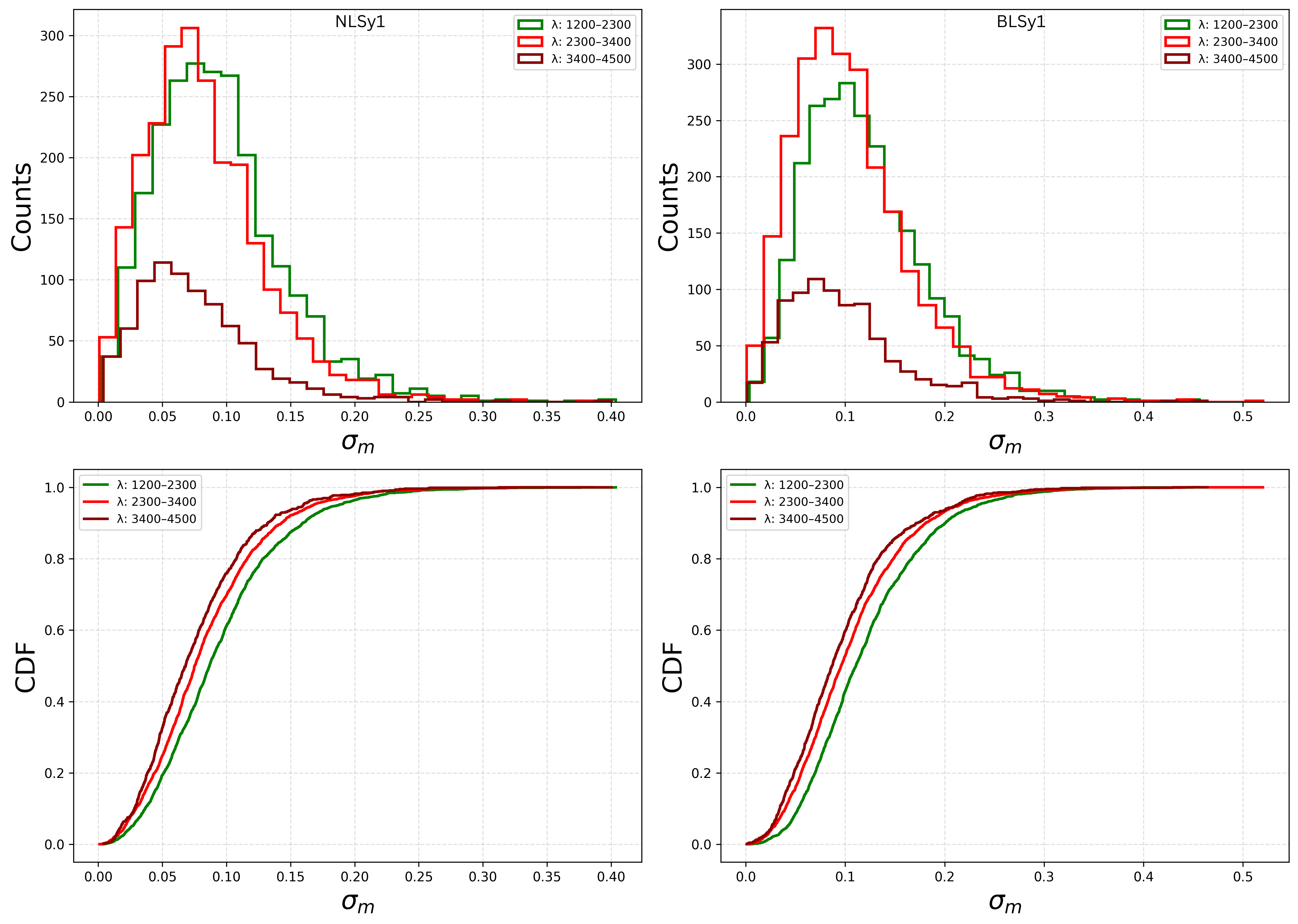}
\caption{Distributions of $\sigma_m$ and CDFs for our sample of NLSy1 galaxies (left panels) and
BLSy1 galaxies (right panels) in the rest frame wavelengths of (i) 1200 $-$ 2300 \AA (green), 
(ii) 2300 $-$ 3400 \AA (red) and (iii) 3400 $-$ 4500 \AA (brown).}
\label{fig-8}
\end{figure*}

The results found here for the high redshift sample of NLSy1 galaxies are similar to that known for the low redshift sample of NLSy1 galaxies \citep{rakshit2017optical,2023Ap&SS.368...68W,2025MNRAS.543..121S}. For example, from an analysis of data in g, r, and i bands from Pan-STARRS, \cite{2023Ap&SS.368...68W} found variability amplitude of 0.142 mag (0.119 mag), 0.137 mag (0.114 mag) and 0.130 mag (0.112 mag) in g, r, and i bands for BLSy1(NLSy1) galaxies. Simlarly, 
\cite{2025MNRAS.543..121S} from ZTF survey found values of 0.374 mag (0.321 mag) and 0.338 mag (0.287 mag) in g and r bands for BLSy1(NLSy1) galaxies. While our results are similar to that found by \cite{2023Ap&SS.368...68W}, they are much lower than that reported by \cite{2025MNRAS.543..121S}.  We note, that the results reported by \cite{2023Ap&SS.368...68W} and \cite{2025MNRAS.543..121S}  
pertain to sources with $z < 0.8$. However, the sample analysed in this work covers a wide redshift range ($0.8 < z < 2.6$), and thus the observed light curves in g, r, and i bands sample different rest wavelengths. We therefore divided the observed effective wavelengths of g, r, and i bands, respectively by (1 + z). The sample thus covers the rest frame wavelengths from $\sim$1250 \AA ~to 4400 \AA. We estimated the amplitude of variability at three rest frame wavelength ranges, namely (i) 1200 $-$ 2300 \AA (short), (ii) 2300 $-$ 3400 \AA (mid) and (iii) 3400 $-$ 4500 \AA (long). 
For NLSy1 galaxies we found mean $\sigma_m$ values of 0.09$\pm$0.05 mag, 0.08$\pm$0.05 mag and 0.08$\pm$0.05 mag respectively for the short, mid and long rest wavelengths. Similarly for BLSy1 galaxies, we found mean values of 0.12$\pm$0.06 mag, 0.10$\pm$0.06 mag and 0.10 $\pm$0.06 mag respectively for the short, mid and long rest wavelengths. Though within error bars, the mean $\sigma_m$ in different rest wavelengths for both BLSy1 and NLSy1 galaxies are similar, KS test indicates that the underlying distributions are indeed different with a clear indication of wavelength dependent variability. 
The histograms and CDFs of the amplitude of variations in different rest frame wavelengths for BLSy1 and NLSy1 galaxies are shown in Fig. $\ref{fig-8}$. We thus found a decreasing trend of variability from shorter to longer rest frame wavelengths. This is also 
evident in the CDFs shown in Fig. $\ref{fig-8}$ and the results of the KS test given in Table \ref{table-5}.

\begin{table*}[!tp]
\small
\caption{Comparison of the amplitudes of variability obtained by DRW and $\sigma_m$ (M refers to the median).}
\label{table-4}
\begin{tabular}{ccccccccc} \hline
       &\multicolumn{4}{c}{NLSy1 galaxies} &\multicolumn{4}{c}{BLSy1 galaxies} \\
Filter &\multicolumn{2}{c}{DRW} &\multicolumn{2}{c}{$\sigma_m$} &\multicolumn{2}{c}{DRW} &\multicolumn{2}{c}{$\sigma_m$} \\
& Average & M & Average & M & Average & M & Average & M \\ \hline
V & 0.19 $\pm$ 0.14 & 0.18 & 0.25 $\pm$ 0.14 & 0.23 & 0.20 $\pm$ 0.13 & 0.20 & 0.27 $\pm$ 0.12 & 0.27 \\
g & 0.13 $\pm$ 0.04 & 0.12 & 0.09 $\pm$ 0.04 & 0.09 & 0.16 $\pm$ 0.07 & 0.15 & 0.13 $\pm$ 0.07 & 0.12 \\
r & 0.11 $\pm$ 0.05 & 0.10 & 0.08 $\pm$ 0.04 & 0.07 & 0.13 $\pm$ 0.06 & 0.12 & 0.10 $\pm$ 0.06 & 0.08 \\
i & 0.10 $\pm$ 0.05 & 0.08 & 0.07 $\pm$ 0.03 & 0.06 & 0.12 $\pm$ 0.09 & 0.11 & 0.09 $\pm$ 0.06 & 0.08 \\ 
\hline
\end{tabular}
\end{table*}

\subsection{Radio sub-sample}
To investigate possible differences in the flux variability characteristics
between radio-detected and radio-undetected sources in our sample
of BLSy1 and NLSy1 galaxies we first identified radio-emitting sources in
our sample. For this we cross-matched our sample with the 
latest version of the Faint Images of the Radio Sky at 
Twenty-Centimeters (FIRST; \citealt{1995ApJ...450..559B}) 
catalog released on 17 December 2014, which contains
a total of 946,432 sources. FIRST is a high resolution radio survey
conducted at 1.4 GHz using the VLA in its B-configuration.
The images from FIRST have an angular resolution of approximately 
5.4 arcsec and a typical 1-$\sigma$ rms noise level of about 0.15 mJy/beam.
Of the 2684 NLSy1 galaxies in our sample, 2262 sources fall within the FIRST 
survey coverage while 422 sources lie outside it.
Thus, of the 2262 sources, 107 are radio-detected, while 2006 are radio-undetected.
Similarly of the 2490 BLSy1 galaxies, 2080 galaxies lie in the FIRST footprint. Among these, 
82 are radio-detected while 1998 are radio-undetected.
The results of the variability analysis of BLSy1 and NLSy1 galaxies separately for radio-detected and radio-undetected sources are given in Table \ref{table-3}. 
Within error bars, the mean $\sigma_m$ values between radio-detected and radio-undetected sources are similar in g, r, and i bands. 

\begin{figure*}[!tp]
\includegraphics[scale=0.1]{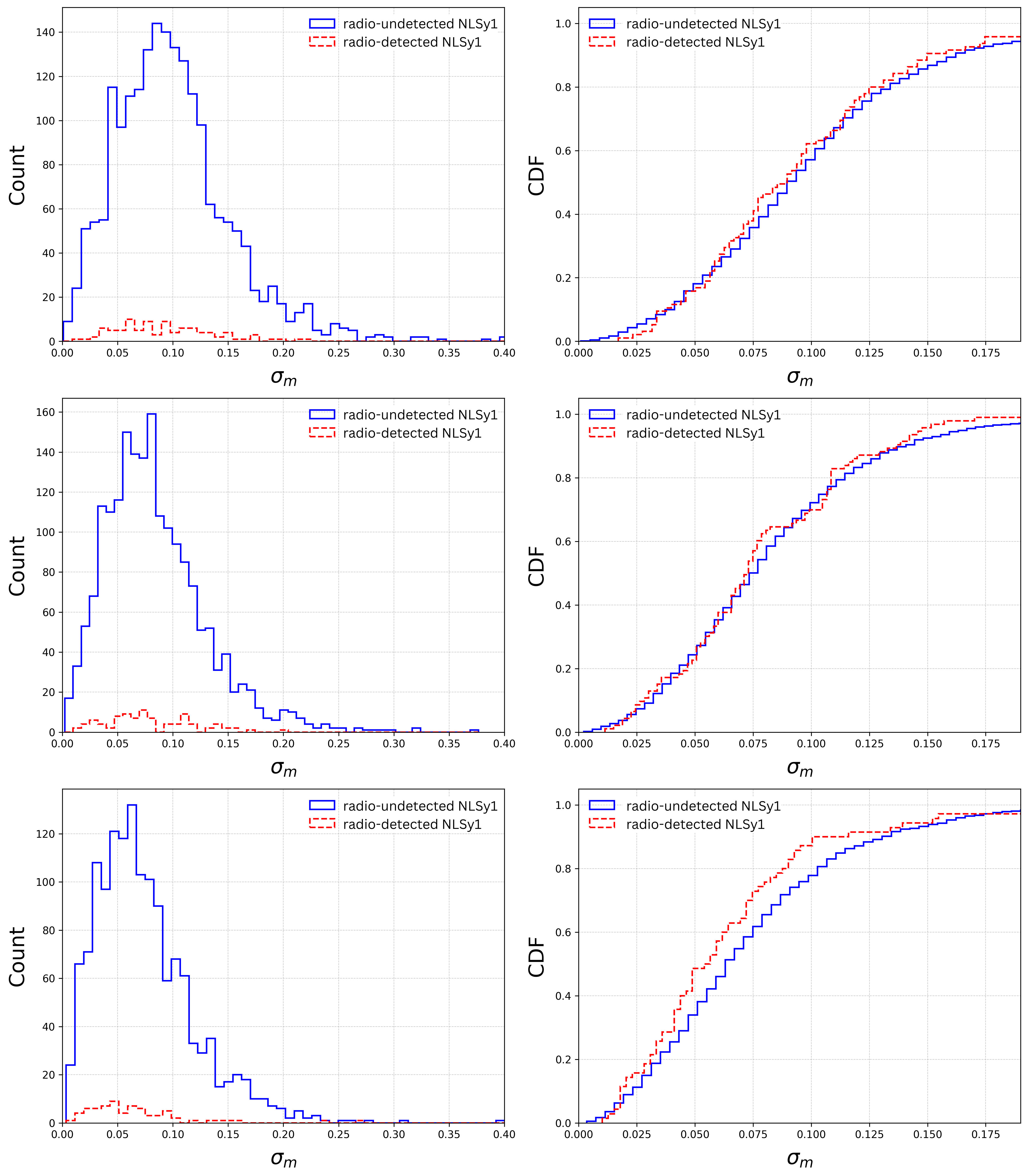}

\caption{Distributions of $\sigma_m$ and CDFs for our sample of NLSy1 galaxies in g band (top panels), r band (middle panels) and i band (bottom panels). Here the solid blue line is for radio-undetected sources, while the red dotted line is for radio-detected} sources.
\label{fig-9}
\end{figure*}

\begin{figure*}[!tp]
\includegraphics[scale=0.1]{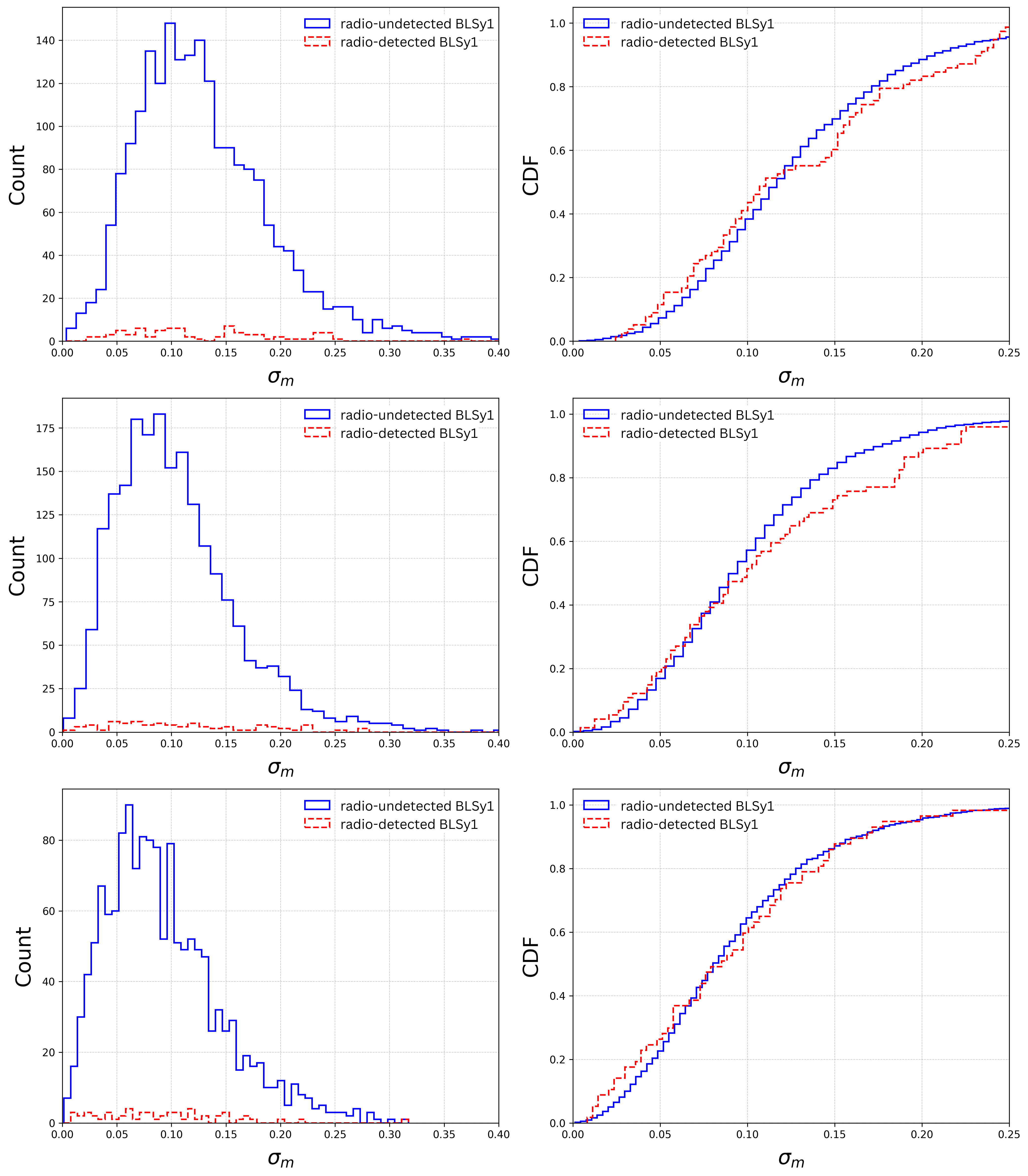}
\caption{Distributions of $\sigma_m$  and CDF for our sample of BLSy1 galaxies in g band (top panels), r band (middle panels) and i band (bottom panels). Here the solid blue line is for radio-undetected sources, while the red dotted line is for radio-detected sources.}
\label{fig-10}
\end{figure*}

\begin{table*}[!tp]
\centering
\small
\caption{Results of KS test. Here, g$_N$, r$_N$ and i$_N$ represent the light curves in g, r, and i bands respectively for the NLSy1 galaxy sample. Similarly,  g$_B$, r$_B$ and i$_B$ denote the  variations in BLSy1 galaxies in g,r and i bands. g$_{N}^{RD}$ r$_{N}^{RD}$ i$_{N}^{RD}$ are the light curves in radio-detected NLSy1 galaxies, while,   g$_{B}^{RD}$ r$_{B}^{RD}$, i$_{B}^{RD}$ refer to the light curves in radio-detected BLSy1 galaxies. Likewise, g$_{N}^{RU}$ r$_{N}^{RU}$ i$_{N}^{RU}$ correspond to the light curves of radio-undetected NLSy1 galaxies and g$_{B}^{RU}$ r$_{B}^{RU}$ i$_{B}^{RU}$ refer to the light curves of the radio-undetected BLSy1 galaxies,\\
NLSy1$_{short}$, NLSy1$_{mid}$, NLSy1$_{long}$ refer to the light curves at short, mid and long rest-frame wavelengths in NLSy1 galaxies, BLSy1$_{short}$, BLSy1$_{mid}$, BLSy1$_{long}$ refer to the light curves at short, mid and long rest-frame wavelengths in BLSy1 galaxies,\\
D is the KS statistic that quantifies the maximum absolute difference between the CDFs of the parameters and P is the probability that the two parameters that are tested are drawn from the same parent distribution.}
\label{table-5}
\begin{tabular}{ccc}
Parameter  & D    & P   \\ \hline
g$_N$ v/s r$_N$               & 0.180 & 0.0792 \\
g$_N$ v/s i$_N$               & 0.339 & 0.0001 \\
r$_N$ v/s i$_N$               & 0.238 & 0.0127 \\
g$_B$ v/s r$_B$               & 0.153 & 0.2963 \\
g$_B$ v/s i$_B$               & 0.296 & 0.0047 \\
r$_B$ v/s i$_B$               & 0.177 & 0.2289 \\
g$_N$ v/s g$_B$               & 0.320 & 0.0002 \\
r$_N$ v/s r$_B$               & 0.265 & 0.0046 \\
i$_N$ v/s i$_B$               & 0.331 & 0.0014 \\
g$_{N}^{RD}$ v/s g$_{N}^{RU}$ & 0.081 & 0.5689 \\
r$_{N}^{RD}$ v/s r$_{N}^{RU}$ & 0.093 & 0.3968 \\ \hline
\end{tabular}%
\hfill
\begin{tabular}{ccc}
Parameter  & D    & P   \\ \hline
i$_{N}^{RD}$ v/s i$_{N}^{RU}$ & 0.146 & 0.1040 \\
g$_{B}^{RD}$ v/s g$_{B}^{RU}$ & 0.114 & 0.2668 \\
r$_{B}^{RD}$ v/s r$_{B}^{RU}$ & 0.141 & 0.1071 \\
i$_{B}^{RD}$ v/s i$_{B}^{RU}$ & 0.083 & 0.8204 \\
NLSy1$_{short}$ v/s NLSy1$_{mid}$  & 0.115 & $<$0.0001 \\
NLSy1$_{short}$ v/s NLSy1$_{long}$ & 0.180 & $<$0.0001 \\
NLSy1$_{mid}$ v/s NLSy1$_{long}$   & 0.091 & $<$0.0001 \\
BLSy1$_{short}$ v/s BLSy1$_{mid}$  & 0.116 & $<$0.0001 \\
BLSy1$_{short}$ v/s BLSy1$_{long}$ & 0.182 & $<$0.0001 \\
BLSy1$_{mid}$ v/s BLSy1$_{long}$   & 0.074 & $<$0.0001 \\ \hline
\end{tabular}
\end{table*}

\begin{figure*}[!tp]
\includegraphics[scale=0.35]{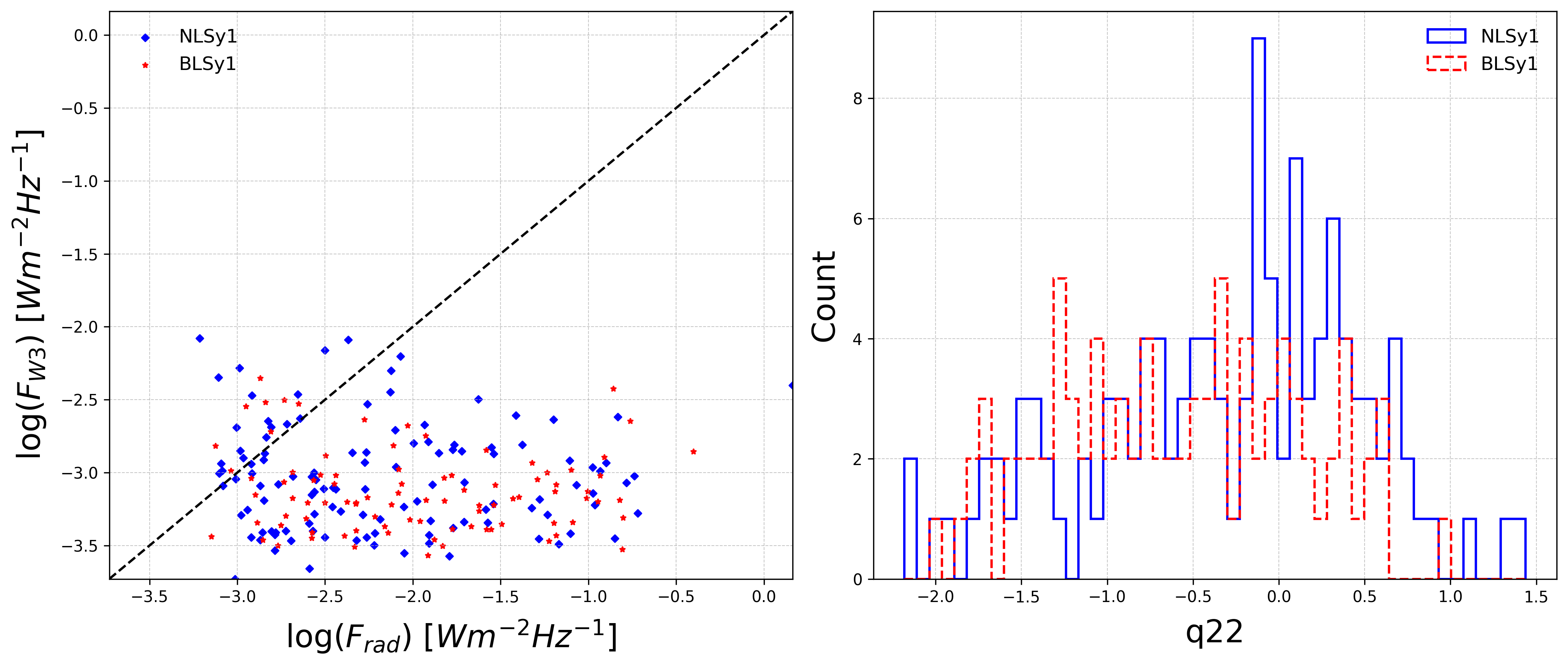}

\caption {Left: WISE W3 flux densities versus 1.4 GHz FIRST flux densities. Here, blue diamonds are NLSy1 galaxies, while red stars are BLSy1 galaxies. The dotted line is the 1:1 line.
Right: Histograms of the q22 parameter for NLSy1 galaxies (solid blue line) and BLSy1 galaxies (dotted red line). }
\label{fig-11}
\end{figure*}

The histograms and CDFs of $\sigma_m$ for the radio-detected and radio-undetected NLSy1 galaxies are shown in Fig. $\ref{fig-9}$. Similarly, in the case of BLSy1 galaxies, the mean $\sigma_m$ values in g, r, and i bands within error bars are similar between radio-detected and radio-undetected sources. Also, both the radio-detected and radio-undetected sources have similar median $\sigma_m$ values in g, r, and i bands. The histograms and CDFs of $\sigma_m$ values in g, r, and i bands are shown in Fig. $\ref{fig-10}$ for BLSy1 galaxies. These results indicate that in both NLSy1 and BLSy1 galaxies, at greater than 90\% confidence level, we found no difference in variability between radio-detected and radio-undetected sources. The results of the KS test are given in Table \ref{table-5}. 
These results are thus in agreement with that found by \cite{2023Ap&SS.368...68W} for 
NLSy1 galaxies with $z$ $<$ 0.8. However, \cite{2017ApJ...842...96R} for the sample 
of NLSy1 galaxies at $z$ $<$ 0.8 found that radio-detected NLSy1 galaxies show
larger $\sigma_m$ than those that are undetected in the radio band. The detected radio emission in our sample of sources, could arise from either 
nuclear (AGN-related) activity or from star formation processes. Given that 
NLSy1 galaxies are known to exhibit enhanced star formation, 
it is possible that their radio emission is dominated by star formation
rather than AGN. To test for the possible cause of radio emission in 
our sample of BLSy1 and NLSy1 galaxies,
we carried out two diagnostics. First, we calculated the q22 parameter defined as \citep{2015MNRAS.451.1795C}

\begin{equation}
q22 = log\left(\frac{F_{22 \mu m}}{F_{1.4 GHz}}\right)
\end{equation}

where $F_{22 \mu m}$ is the flux density in the WISE $W4$ band (22 $\mu$m) and F$_{1.4 GHz}$ is the radio flux density at 1.4 GHz from FIRST. The distributions of q22 for our sample of sources are shown in Fig. $\ref{fig-11}$. We found that approximately 85\% of NLSy1 galaxies have q22 $<$ 1 suggesting that their radio emission is AGN  dominated. Similarly, about 90\% of BLSy1 galaxies  have q22 $<$ 1, indicating that their radio emission is likely powered by AGN activity, possibly due to relativistic jets. Second, we plotted our sources in the flux density at $W3$ band of WISE against the 1.4 GHz flux density from FIRST.

Also, shown in the same figure is the 1:1 line. Sources above the line are considered 
star formation dominated while those below the line are AGN dominated \citep{2021ApJ...910...64K}. In this diagram, about 97\% of NLSy1 galaxies lie below the line and  all BLSy1 galaxies lie below the line indicating that the radio emission in them is primarily driven by AGN. From both diagnostics it is evident that the majority of sources in our sample exhibit radio emission that is likely to be AGN  dominated, possibly due to the presence of relativistic jets.  However, the observed similarity in the variability amplitudes between radio-detected and radio-undetected sources in our sample of NLSy1 and BLSy1 galaxies suggests that relativistic jets have a negligible role in the observed long-term optical variability in our sample. We however, note the number of radio-detected sources is relatively small. This is a limitation of this work, and similar analysis on a large number of radio-detected and radio-undetected sources is needed to constrain the role of disk-driven and jet-driven mechanisms for the long-term optical variability of NLSy1 and BLSy1 galaxies.

\FloatBarrier

\subsection{Colour variations}
\begin{figure*}[!tp]
\vbox{
     \hbox{
          \includegraphics[scale=0.32]{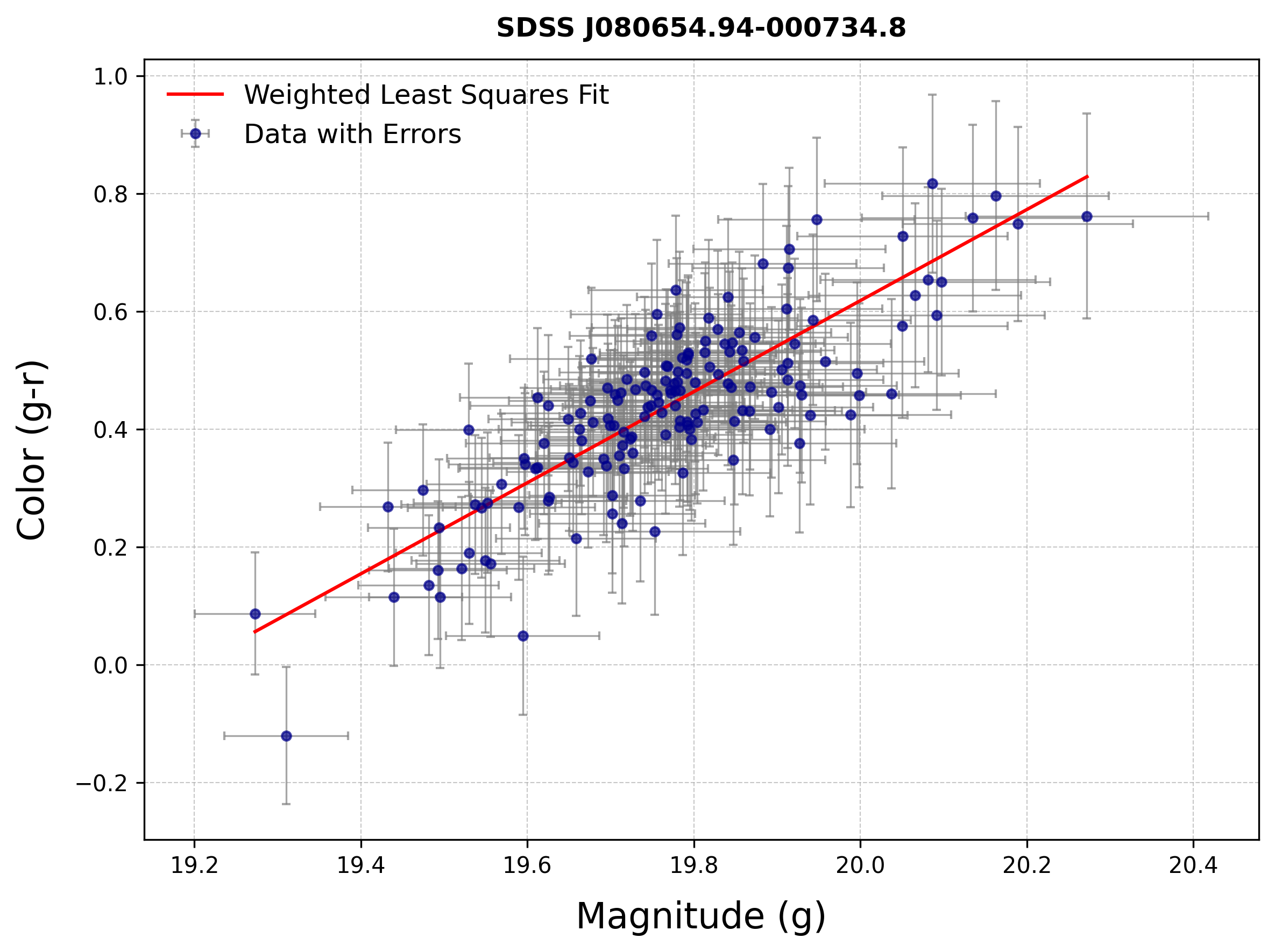}
          \includegraphics[scale=0.32]{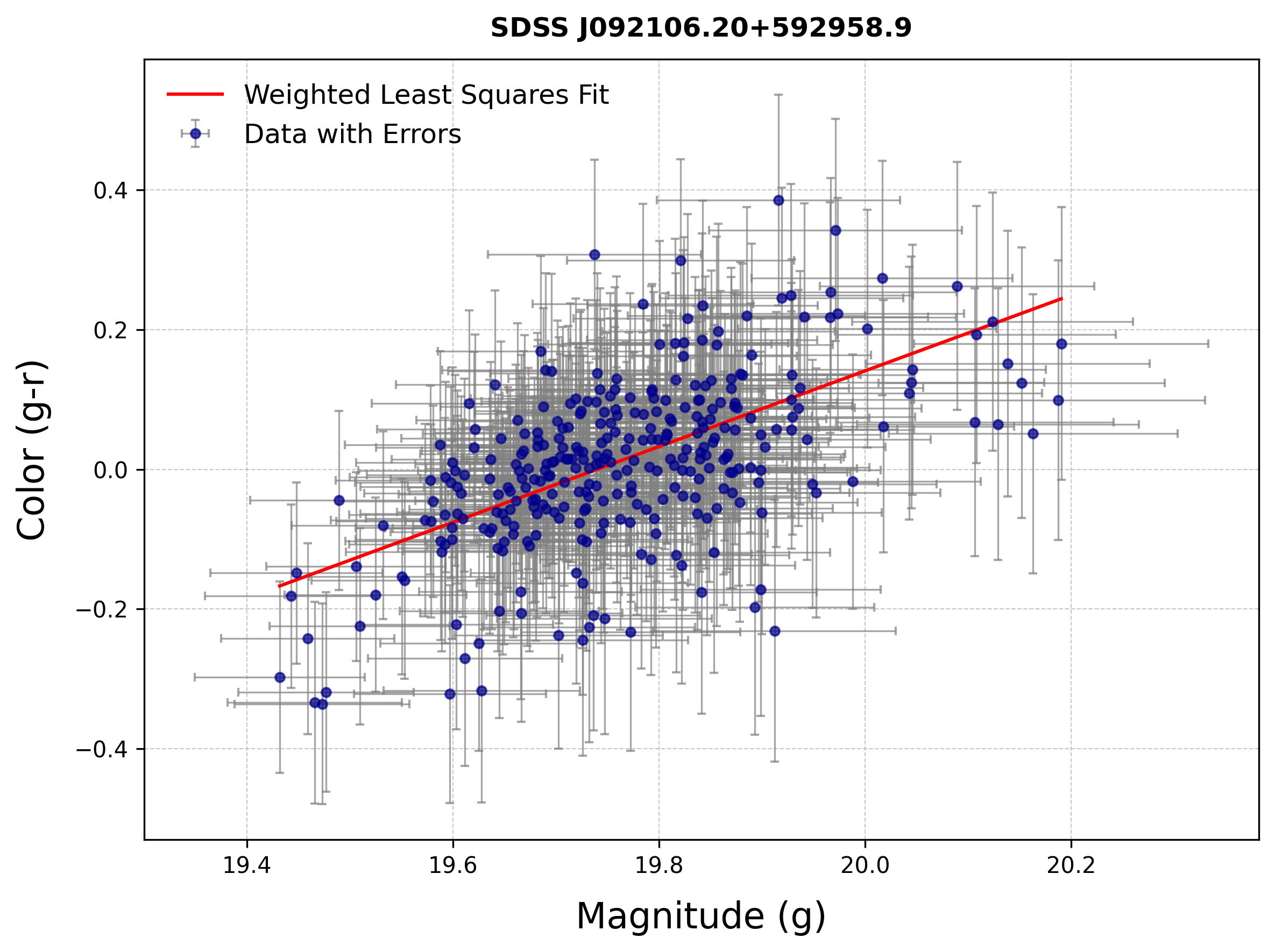}

          }
     \hbox{
          \includegraphics[scale=0.32]{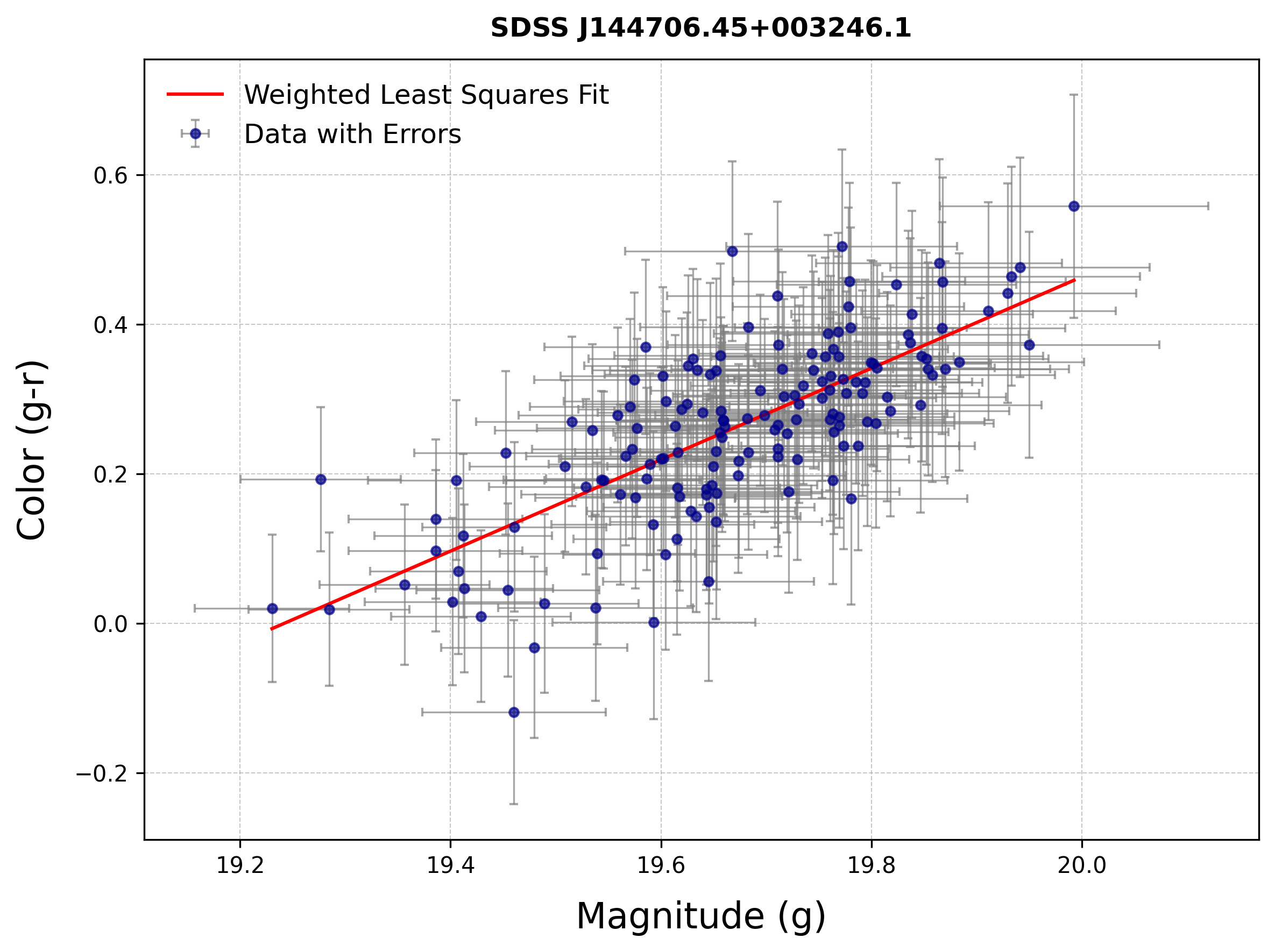}
          \includegraphics[scale=0.32]{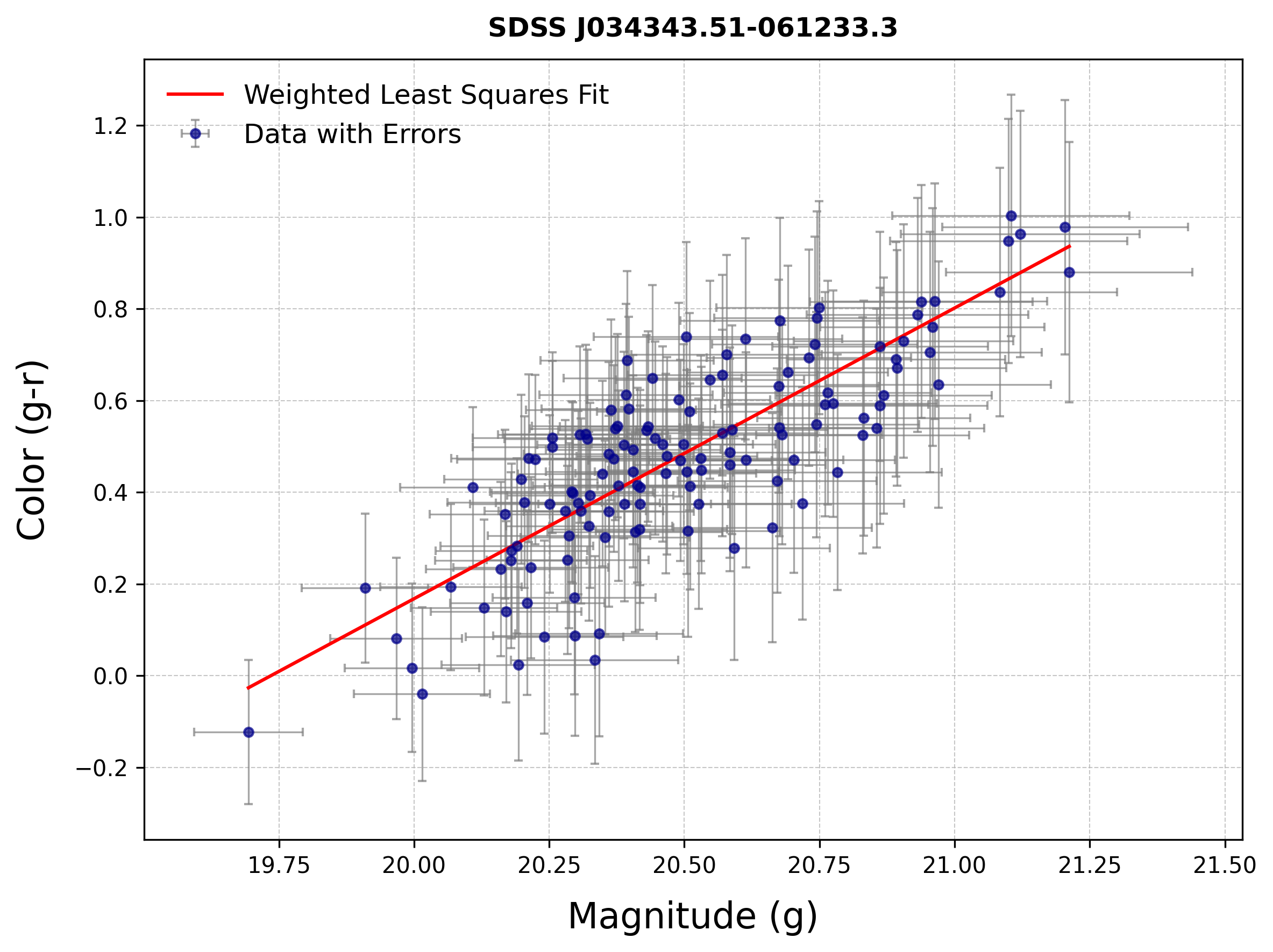}
          }
     \hbox{
          \includegraphics[scale=0.32]{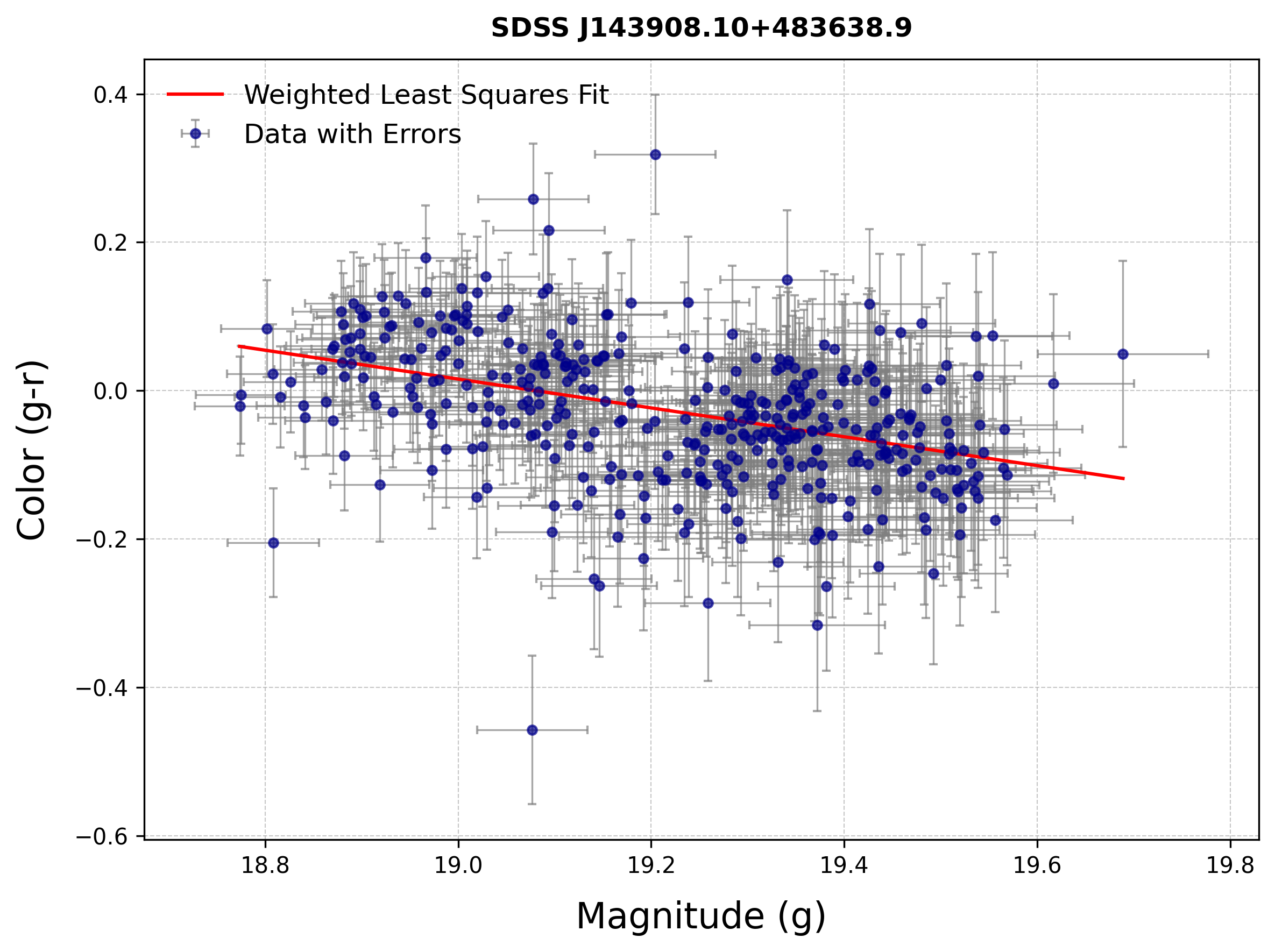}
          \includegraphics[scale=0.32]{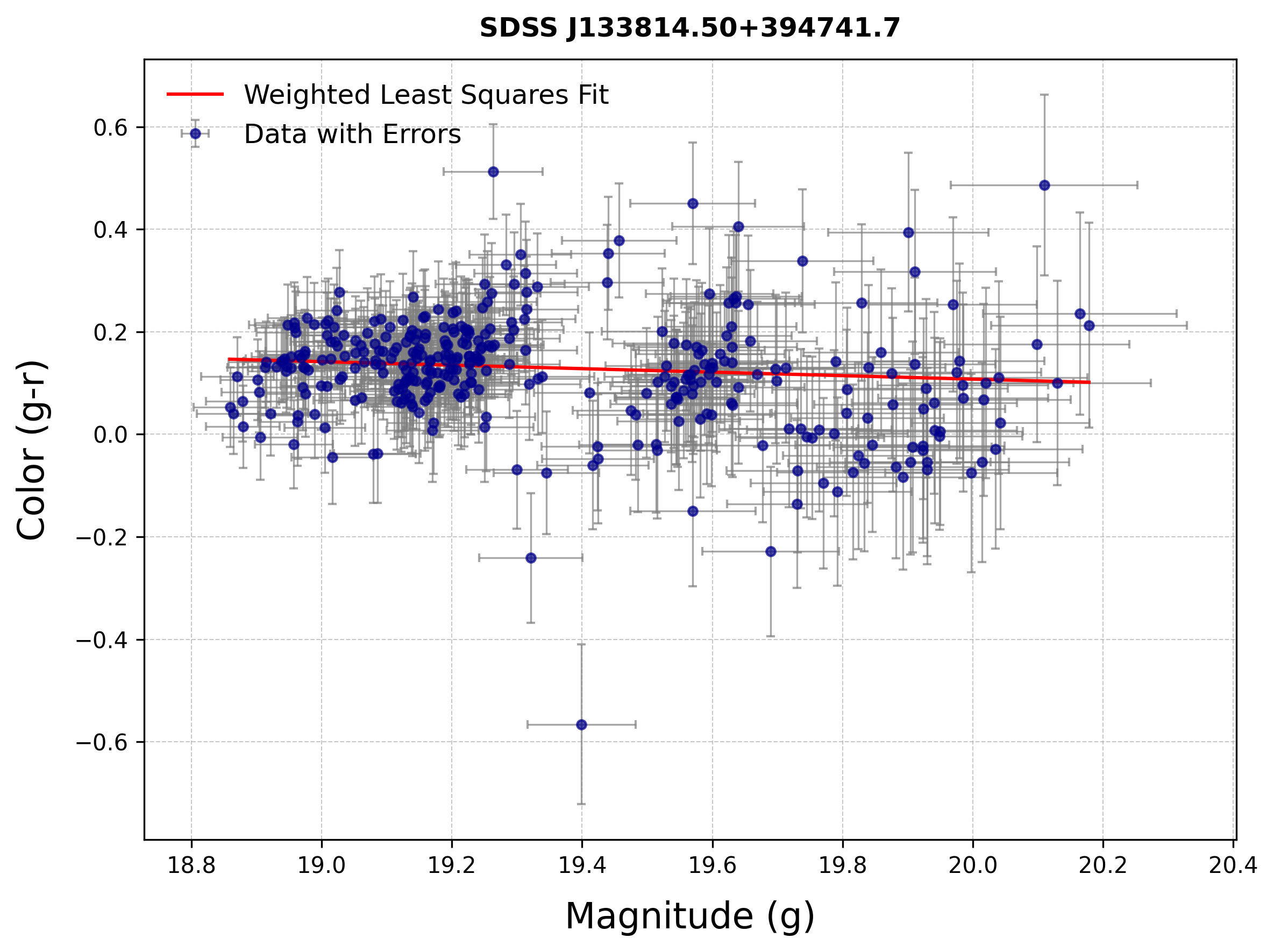}

          }
       }
\caption{Example colour (g$-$r) - magnitude (g band) diagrams for NLSy1 galaxies (left panels) and BLSy1 galaxies (right panels). The names of the sources are given at the top of each panel.}
\label{fig-12}
\end{figure*}

We investigated the colour-variability of our sample of NLSy1 and BLSy1 galaxies on day-like timescales. This was done by generating colour-magnitude diagrams which are useful in probing the underlying physical processes responsible for optical flux variations. For a large fraction of sources we found a bluer when brighter (BWB) trend. Specifically, approximately 90\% of both NLSy1 and BLSy1 galaxies in our sample showed a BWB trend, while the remaining 10\% showed no significant correlation between optical colour and brightness. Notably, only one source in each of our sample of NLSy1 and BLSy1 galaxies showed a redder when brighter (RWB) trend. Example colour variations are shown in Fig. $\ref{fig-12}$ and the results of the colour variability analysis are given in Table \ref{table-6}.  

 Colour variations in AGN are commonly quantified by fitting the data in colour-magnitude space \citep{1999MNRAS.306..637G,2004ApJ...601..692V,2005ApJ...633..638W}, as done here. However, according to \cite{2012ApJ...744..147S}, fitting in colour-magnitude space suffers from co-variances between uncertainties in colour and magnitude, which can  lead to inaccurate results on colour variability. To mitigate this issue, \cite{2012ApJ...744..147S} proposed fitting the variability in magnitude-magnitude space and then transforming the results into colour-magnitude relations. Following \cite{2012ApJ...744..147S} we examined the colour variability nature of our sample of BLSy1 and NLSy1 galaxies by fitting the g-band magnitudes against r-band magnitudes. The slope of the magnitude-magnitude relation was then used to quantify colour variability. To remove redshift dependent biases in the colour variability behaviour we applied the correction described by
\cite{2012ApJ...744..147S}. We define

\begin{equation}
S_{gr}^{corr} = S_{gr} + K(z)
\end{equation}

Here, $S_{gr}$ is the colour index, defined as

\begin{equation}
S_{gr} = (S_{gr}^{\prime} - 1 )
\end{equation}

\begin{figure*}[!tp]
\includegraphics[scale=0.5]{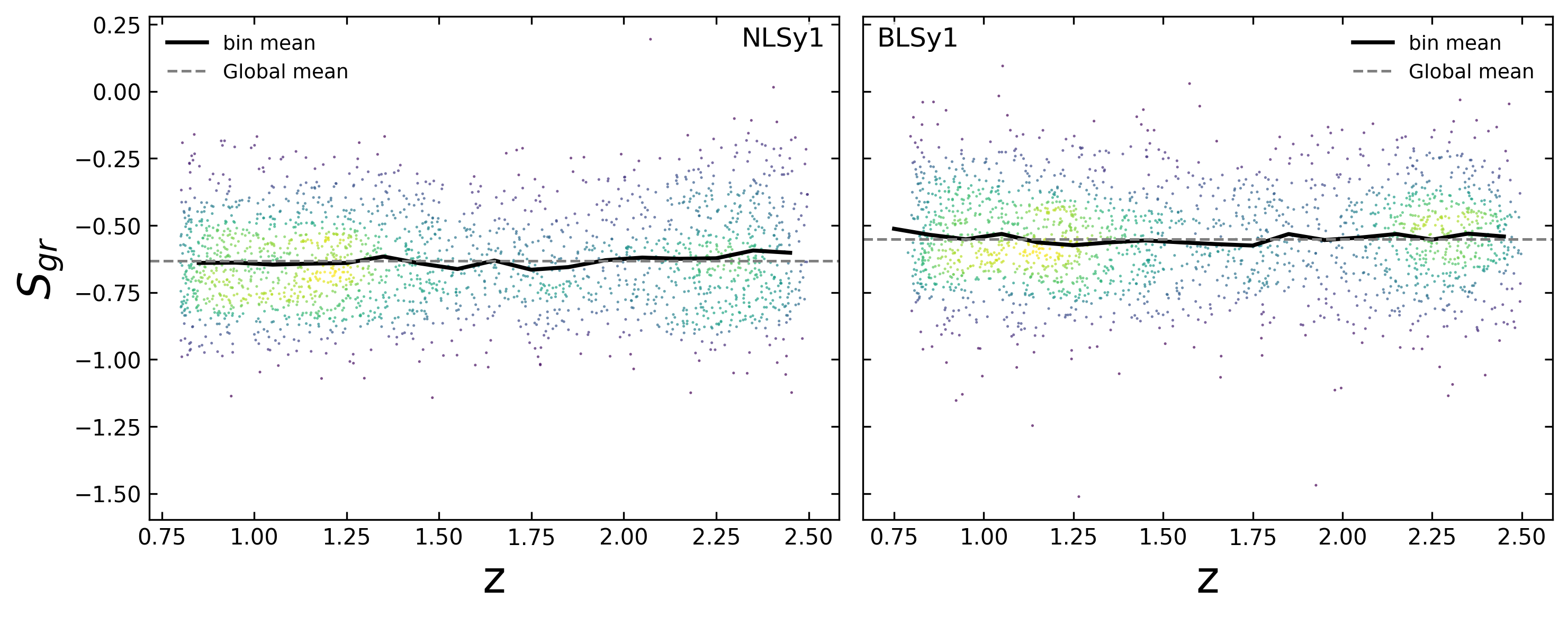}
\caption{Distribution of the colour variability index S$_{gr}$ as a function of redshift for NLSy1 galaxies (left panel) and BLSy1 galaxies (right panel). Here the dashed lines represent the mean S$_{gr}$ for the whole sample and the solid lines represent the mean S$_{gr}$ at redshift intervals of $\Delta z = 0.1$.}
\label{fig-13}
\end{figure*}

$S_{gr}^{\prime}$ is the slope of the linear fit to g- and r-band magnitudes, K(z) represents the redshift dependent deviation computed as the difference between the mean values of S$_{gr}$ of the full sample and the mean of S$_{gr}$ within redshift bins of width $\Delta z$ =  0.1. This is shown in Fig. $\ref{fig-13}$ for both NLSy1 and BLSy1 galaxies.

A value  $S_{gr}^{corr}$ $< 0$ corresponds to a BWB trend, while $S_{gr}^{corr}$ $> 0$ corresponds to a RWB trend. If $S_{gr}^{corr}$ $\sim 0$ (within errors), the source is considered to show no colour variability. Using this approach too, we found a majority of sources in both the samples of BLSy1 (83\%) and NLSy1 (84\%) galaxies to show a BWB trend. This is slightly larger than the value of about 74\% and 79\% found for the low redshift sample ($z < 0.8$) of NLSy1 and BLSy1 galaxies \citep{2025MNRAS.543..121S}. The details are given in Table \ref{table-6}.

\begin{table}[!htbp]
\caption{Results on the analysis of colour (g$-$r) variability against g-band brightness.
The values quoted in parentheses pertain to colour variations from fits to data in 
magnitude-magnitude plane.}
\label{table-6}
\begin{tabular}{ccc} \hline
Parameter & NLSy1 galaxies & BLSy1 galaxies \\ \hline
Total     & 2341 (2490)    & 2312 (2490)    \\
BWB       & 2106 (2081)    & 2070 (2056)    \\
RWB       & 1    (1)       & 1    (2)      \\
No trend  & 234  (408)     & 241  (432)    \\ \hline
\end{tabular}
\end{table}

The observed prevalence of BWB trend in our sample of sources can be understood due to flux variations driven by localised temperature fluctuations rather than changes in the accretion rate. Such localised temperature fluctuations can lead to increased emission at shorter wavelengths when the source brightens thereby leading to a BWB trend \citep{2012ApJ...744..147S}. Such BWB trend as noticed here for the high redshift sample is also observed by other authors. For instance, using photometric data from ZTF, 
\cite{2025MNRAS.543..121S} found that the majority of NLSy1 and BLSy1 galaxies at $z < 0.8$, also exhibited a BWB trend. In the mid-infrared too, NLSy1 galaxies are known to show a BWB trend \citep{2019MNRAS.483.2362R}. The BWB trend found in this work for a  majority of BLSy1 and NLSy1 galaxies is consistent with what is known for BLSy1 and NLSy1 galaxies at $z < 0.8$.

\subsection{Lag between different optical bands}
\begin{figure*}[!tp]
\hbox{
      \includegraphics[scale=0.17]{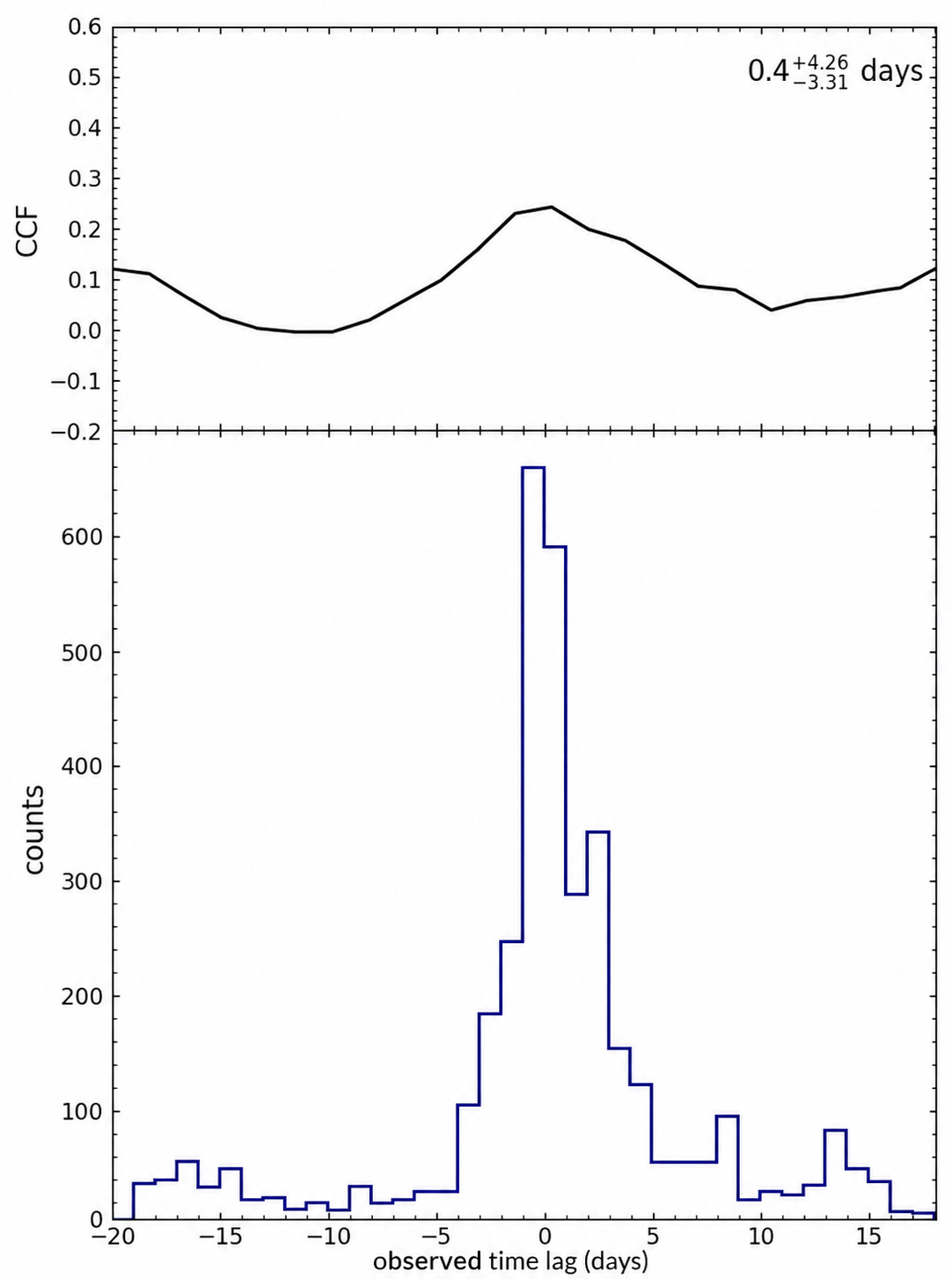}
      \includegraphics[scale=0.17]{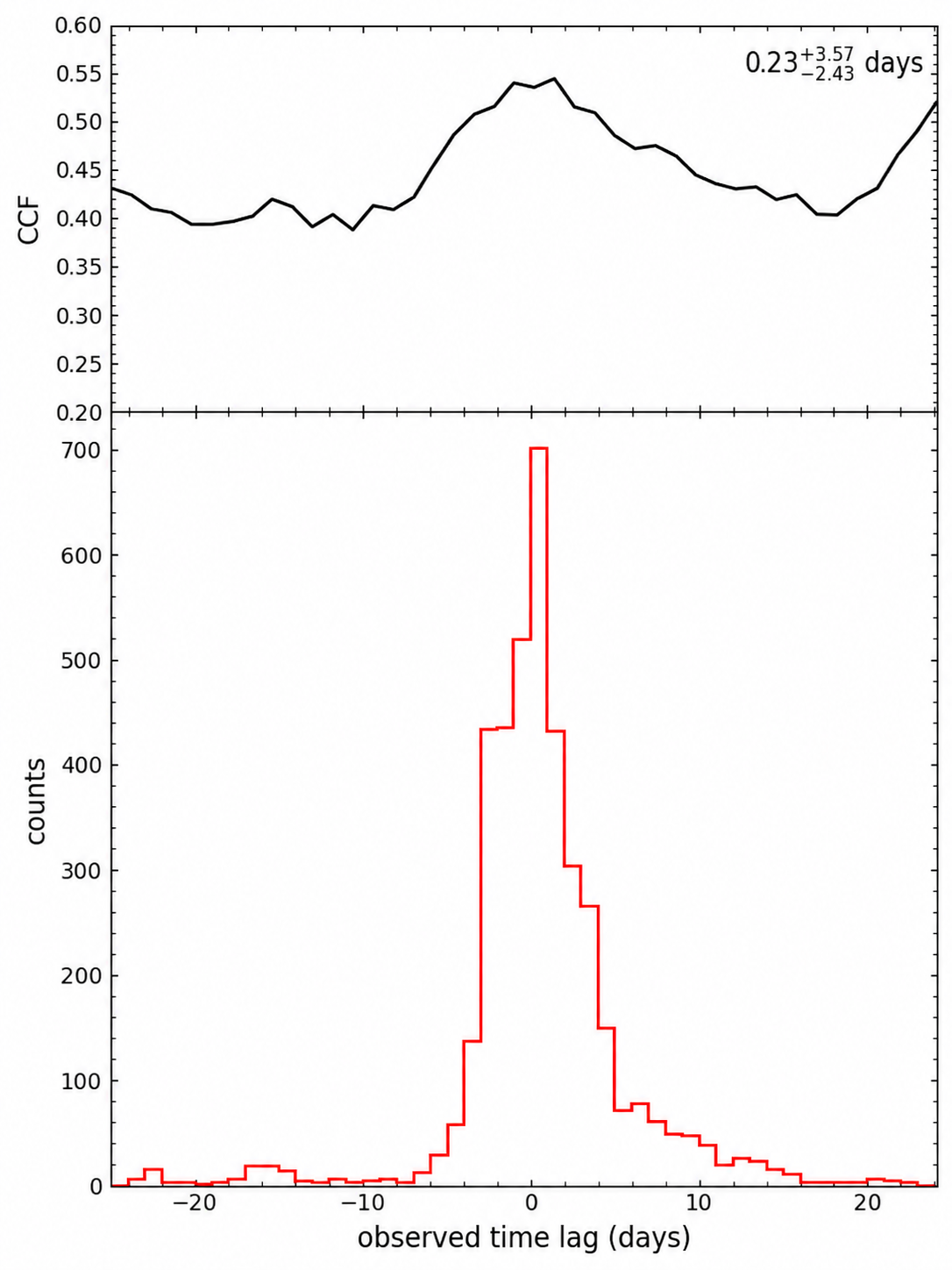}

     }
\caption{The cross-correlation functions for a NLSy1 galaxy (left panels) and a BLSy1 galaxy (right panels). The top panels refer to the cross-correlation function between g and r bands, obtained using ICCF. The bottom panels show the distribution of the centroids of the CCFs obtained using 5000 Monte Carlo simulations. The value of the centroid of the distribution along with the associated error is given in the corresponding panel. The values are consistent with zero lag.}
\label{fig-14}
\end{figure*}

To compute the time lag between g and \r bands, we used the Interpolated Cross-Correlation Function (ICCF; \citealt{1986ApJ...305..175G,1987ApJS...65....1G}) for which we used its python based implementation namely pyCCF \citep{2018ascl.soft05032S}. To find the uncertainty in the derived lags, we followed a Monte Carlo (MC) approach by incorporating flux randomization and  random subset selection following the procedure outlined in \cite{1998PASP..110..660P}. For each MC iteration, we generated the cross-correlation function (CCF) and for each CCF, we estimated the lag using the centroid of the CCF, considering all points that are within 80\% of the peak of the CCF. This was repeated 5000 times for each source and for each MC iteration, we computed the CCF and calculated the centroid of the CCF. The mean of the distribution of centroids of the CCFs is considered as the lag between the g- and r-band light curves and the uncertainties correspond to the 15.87$^{th}$ and 84.13$^{th}$ percentiles of the posterior distribution of the CCF centroids, which corresponds to 1-$\sigma$ error in the case of a Gaussian distribution. We found no significantly detected time lag between g and r bands, longer than the time resolution of the observations used in this work. The CCF for one NLSy1 galaxy and a BLSy1 galaxy is shown in Fig. $\ref{fig-14}$. Also shown in Fig. $\ref{fig-14}$ are the distributions of the centroids of CCFs from MC simulations for the sources in the sample of NLSy1 and BLSy1 galaxies.

\subsection{Relation between variability amplitude and physical parameters}
Previous studies have shown that AGN variability can be linked to several physical parameters such as black hole mass, Eddington ratio and radio jet power. However, the exact physical mechanisms driving this variability remain unclear. It has been shown by \citet{kawaguchi2000temporal} that certain types of events that produce a power-law power density spectrum  may originate from instabilities 
within the accretion disk. In the case of jetted AGN, shock waves propagating through the jet can also contribute to variations in the light curves. Given this background, we tried to investigate how key physical parameters such as the black hole mass, the Eddington ratio, and radio jet power for the radio-detected sources (considering the observed radio emission is due 
to relativistic jets) could play a role in the observed variability. The Eddington ratio ($\lambda_{Edd}$) is defined as $\lambda_{Edd} = \frac{L_{Bol}}{L_{Edd}}$, where $L_{Bol}$ is the bolometric luminosity determined by $L_{Bol} = 9 \times \lambda L_{\lambda}(5000 \AA)$ erg s$^{-1}$ and $L_{Edd}$ is the Eddington luminosity defined as $L_{Edd} = 1.3 \times 10^{31}$ (M$_{BH}$/M$_{\odot}$) erg s$^{-1}$.

%nlsy1 corre
\begin{figure*}[!tp]
\includegraphics[scale=0.34]{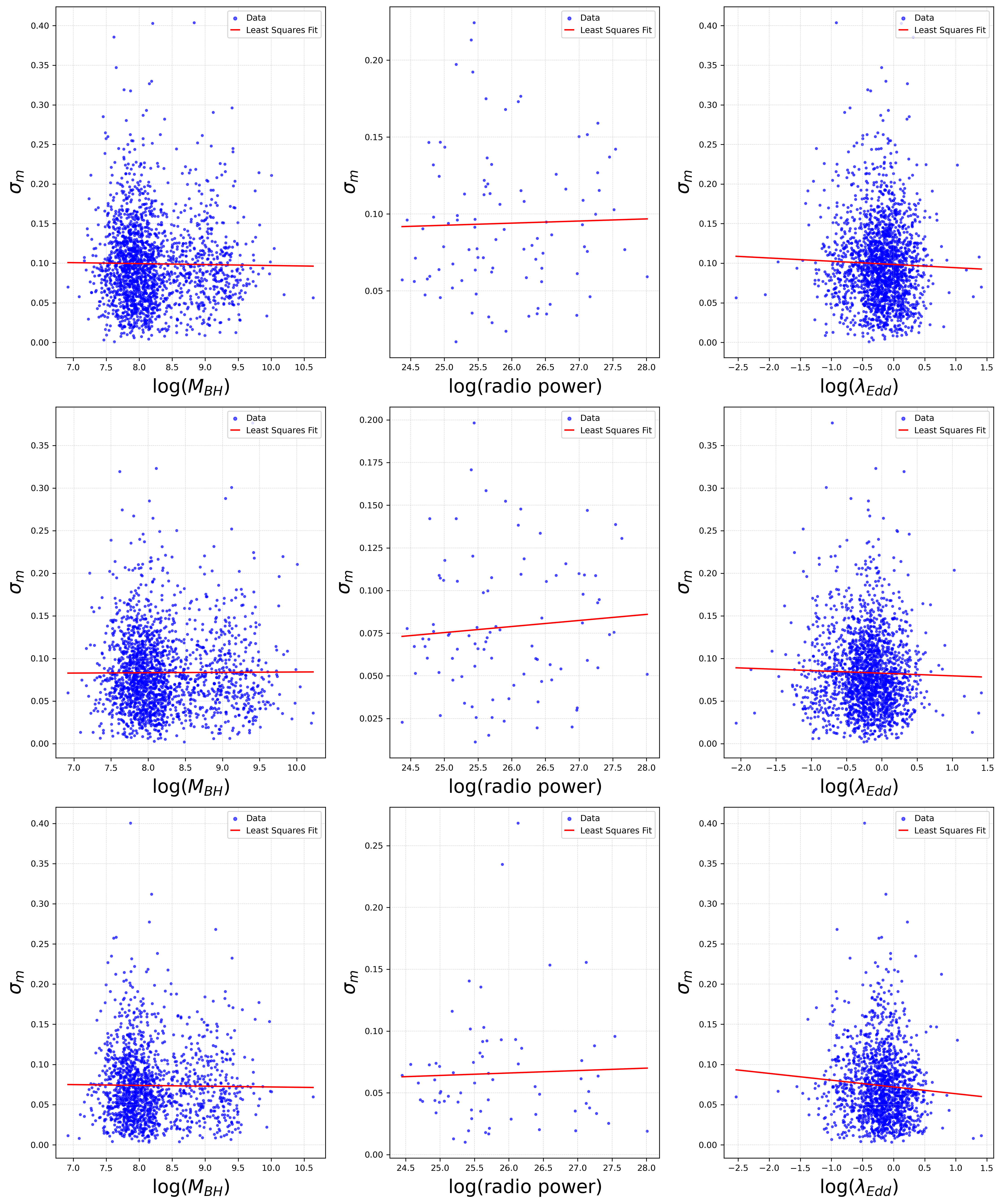}
\caption{Correlation between $\sigma_m$ and M$_{BH}$ (left panels),  $\sigma_m$ and radio power (middle panels) and $\sigma_m$ and $\lambda_{Edd}$ (right panels) for the NLSy1 galaxy sample. From top to bottom, the panels refer to g, r, and i bands respectively.  In all the panels the solid lines are the linear least squares fit to the data.}
\label{fig-15}
\end{figure*}

%blsy1 correlation
\begin{figure*}[!tp]
\includegraphics[scale=0.34]{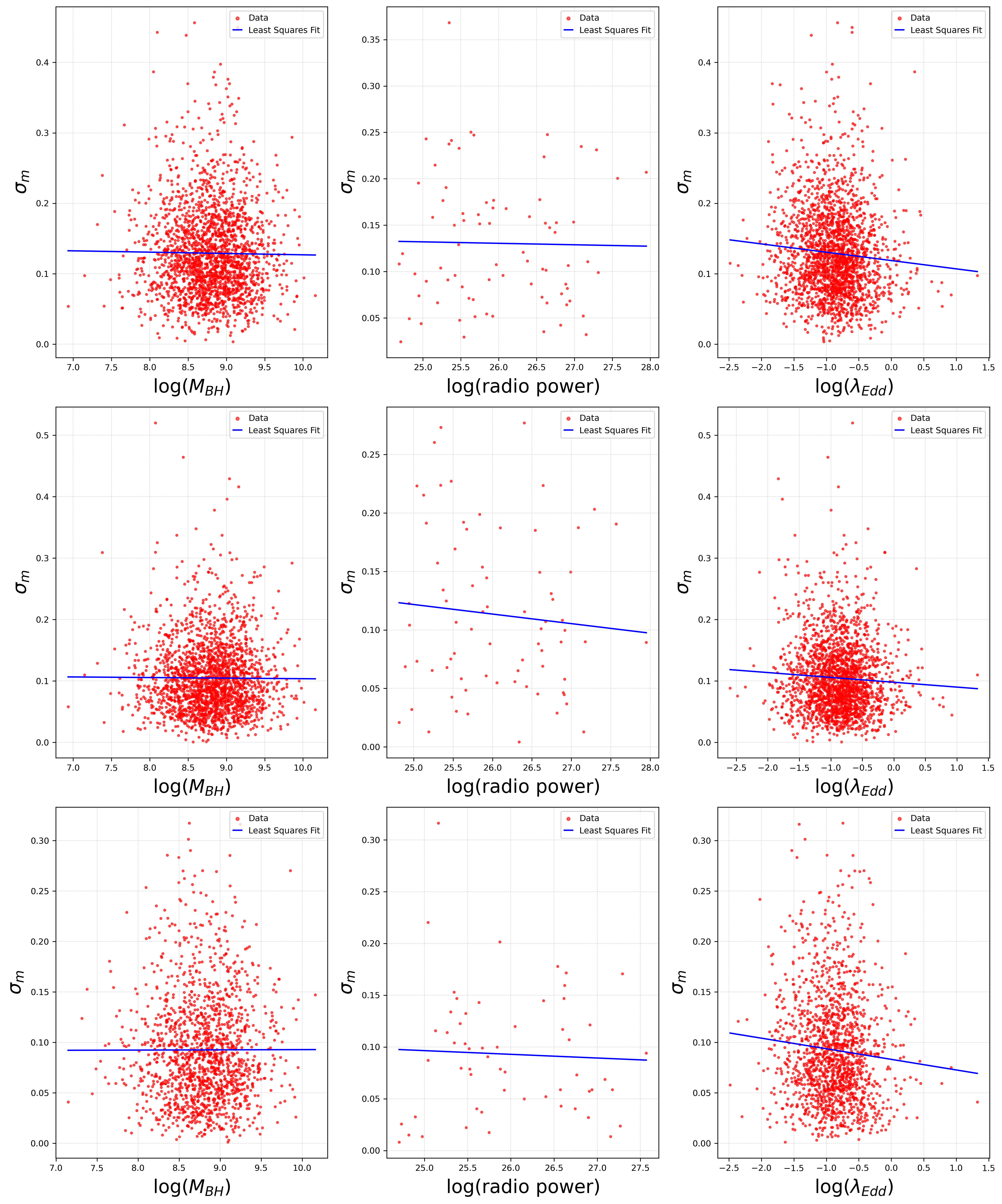}
\caption{Correlation between $\sigma_m$ and M$_{BH}$ (left panels),  $\sigma_m$ 
and radio power (middle panels) and $\sigma_m$ and $\lambda_{Edd}$ (right panels) for the BLSy1 galaxy sample. From top to bottom, the panels refer to g, r, and i bands respectively. In all the panels the solid lines are the linear least squares fit to the data.}
\label{fig-16}
\end{figure*}

%corr table
\begin{table*}[!tp]
\centering
\small
\caption{Results of the correlation analysis between $\sigma_m$ and M$_{BH}$, $\lambda_{Edd}$ and radio power for BLSy1 and NLSy1 galaxies. Here R is the linear correlation coefficient and P is the probability for no correlation. Values with P $<$ 1 $\times$ 10$^{-3}$ are entered as 0.00}
\label{table-7}
\begin{tabular}{ccrcrc} \hline
Parameter & Band     & \multicolumn{2}{c}{NLSy1 sample} & \multicolumn{2}{c}{BLSy1 sample} \\ 
                                 &     &  R       &  P      & R          & P \\ \hline
$\sigma_m$ v/s M$_{BH}$          & g   & $-$0.005 & 0.851   & $-$0.012   & 0.563   \\
                                 & r   &    0.024 & 0.358   & $-$0.007   & 0.750   \\
                                 & i   & $-$0.013 & 0.640   &    0.002   & 0.945   \\
$\sigma_m$ v/s $\lambda_{Edd}$   & g   & $-$0.034 & 0.182   & $-$0.080   & 0.000   \\ 
                                 & r   & $-$0.046 & 0.084   & $-$0.057   & 0.008   \\ 
                                 & i   & $-$0.067 & 0.016   & $-$0.083   & 0.001   \\ 
$\sigma_m$ v/s Radio power       & g   &    0.034 & 0.758   & $-$0.018   & 0.876   \\ 
                                 & r   &    0.080 & 0.475   & $-$0.089   & 0.450   \\ 
                                 & i   &    0.050 & 0.701   & $-$0.046   & 0.736   \\  \hline

\end{tabular}
\end{table*}

For analysis of the correlation if any between the amplitudes of variability and the various physical properties of the sources, we decided not to use the results from the analysis of V-band data from CRTS. This is because, the photometric errors in CRTS light curves are underestimated \citep{2017MNRAS.472.4870S}.
The correlations between $\sigma_m$ against M$_{BH}$, $\lambda_{Edd}$ and radio power in g, r and i bands for NLSy1 and BLSy1 galaxies are given in Fig. $\ref{fig-15}$ and Fig. $\ref{fig-16}$  respectively.
Shown on the figures in solid lines are the linear least squares fit to the data. To quantify the relationships, we carried out Pearson test between $\sigma_m$ and
the physical properties of the sources such as M$_{BH}$, $\lambda_{Edd}$ and radio power. The results of the correlation analysis are given in Table \ref{table-7}. We found no correlation between $\sigma_m$ and M$_{BH}$ in g, r, and i bands in both NLSy1 and BLSy1 galaxies. 
This is in contrast to that found by \cite{2017ApJ...842...96R} wherein, the authors found a positive correlation between variability and M$_{BH}$. 
Similarly we found no correlation between $\sigma_m$ and radio power. This is in agreement with that found by \cite{2023Ap&SS.368...68W}, however, in contrast to that found by \cite{2017ApJ...842...96R}. We found anti-correlation between $\sigma_m$ and $\lambda_{Edd}$ in g, r, and i bands in the case of BLSy1 galaxies,
while such an anti-correlation between $\sigma_m$ and $\lambda_{Edd}$ was observed only in the r- and i-band observations of NLSy1 galaxies at greater than 90\% confidence. Such a negative correlation between $\sigma_m$ and $\lambda_{Edd}$ in NLSy1 and BLSy1 galaxies has been reported by \cite{2017ApJ...842...96R} and well understood in the scenario of the standard accretion disk \citep{1973A&A....24..337S}. The observed long-term variability characteristics in g, r, and i bands in NLSy1 and BLSy1 galaxies and their observed correlation with M$_{BH}$, radio-power and $\lambda_{Edd}$ could be due to variations in the accretion disk \citep{2025ApJ...992..130Y}.

\FloatBarrier
\newpage
\section{Conclusions}
We carried out an investigation of the long-term optical variability characteristics
of a sample of 2490 high redshift  NLSy1 galaxies at $z$ $>$ 0.8. As a control sample
we also analysed a sample of 2490 BLSy1 galaxies. The results of the work are 
summarized below

\begin{enumerate}
\item NLSy1 galaxies tend to show lower amplitude of variability compared
to BLSy1 galaxies in all the optical wavebands (V, g, r, and i) analysed in this
work.
\item We found a clear wavelength dependent variability with the 
amplitude of variability increasing towards shorter wavelengths. This is 
seen in both NLSy1 and BLSy1 galaxies.
\item We found no difference in the amplitude of flux variations (in different optical
bands) between both radio-detected and radio-undetected sources in our sample
of BLSy1 and NLSy1 galaxies.
\item On spectral variations, about 90\% of the sources in BLSy1 and NLSy1 galaxies
showed a BWB trend. No significant spectral variations could be ascertained in
10\% of our sources in NLSy1 and  BLSy1 samples. 
\item We found no time lag between flux variations in g and r bands.
\item We found no correlation between $\sigma_m$ and M$_{BH}$ in g, r, and i bands in both NLSy1 and BLSy1 galaxies. We found anti-correlation between $\sigma_m$ and $\lambda_{Edd}$ in g, r, and i bands at greater than 99\% confidence in our sample of BLSy1 galaxies. However, in NLSy1 galaxies, such negative correlation between $\sigma_m$ and $\lambda_{Edd}$ is seen in r- and i-band observations.
\end{enumerate}

\bmhead{Acknowledgements}
We are very thankful to the referee for sharing his/her detailed comments with us. The comments have enriched the contents of this work.

\section*{Declarations}
\noindent{\textbf{Funding:}}
The authors declare that no funds were received during the preparation of this manuscript.\\

\noindent{\textbf{Competing interests:}}
The authors declare no competing interests.\\

\noindent{\textbf{Ethics approval and consent to participate:}}
Not applicable\\

\noindent{\textbf{Consent for publication:}}
All the authors consented to the publication of this work.\\

\noindent{\textbf{Data availability:}}
All the data analysed in this work are available in \url{https://nesssi.cacr.caltech.edu/CRTS/} (CRTS) and \url{https://irsa.ipac.caltech.edu/cgi-bin/Gator/nph-scanmission=irsa\&submit=Select\&projshort=ZTF} (ZTF)\\

\noindent{\textbf{Materials availability:}}
Not applicable.\\

\noindent{\textbf{Code availability:}}
All codes used in this work were developed by the authors using Python.\\

\noindent{\textbf{Author contribution:}}
All authors contributed to the study from the initial conception to the final arrival of the manuscript. AD and AS collected and analysed the data. All authors contributed to the text. CSS and SR provided detailed comments, which shaped the structure of the manuscript.\\

\bibliography{references1} % common bib file

\end{document}